\documentclass[11pt,nofootinbib,preprint]{revtex4}
\usepackage{mathrsfs}
\usepackage{graphicx}
\usepackage{amsmath}
\usepackage{amsfonts}
\usepackage{amssymb}
\usepackage{color}

\usepackage{epsfig}
\usepackage{CJK}
\usepackage{graphicx}
\usepackage{epsfig}
\usepackage{eepic}
\usepackage{bbm}
\usepackage{dcolumn}
\usepackage{bm}
\usepackage{ulem}
\usepackage{slashbox}
\usepackage{multirow}
\usepackage{slashed}

\newcommand{\omits}[1]{}

\def\bc{\begin{center}}
\def\nno{\nonumber}
\def\ec{\end{center}}
\def\be{\begin{eqnarray}}
\def\ee{\end{eqnarray}}

\definecolor{dyellow}{rgb}{1.,0.8,.0}
\definecolor{myblue}{rgb}{.1,.1,.7}
\definecolor{dcyan}{rgb}{.0,.6,.6}
\definecolor{cyan}{rgb}{0.4,1.0,1.0}
\definecolor{dmagenta}{rgb}{0.6,0.0,0.6}
\definecolor{brown}{rgb}{0.6,0.2,0.}
\definecolor{darkblue}{rgb}{.0,.0,0.5}
\definecolor{darkred}{rgb}{0.75,0.0,0.0}
\definecolor{orange}{rgb}{1.,.6,.0}
\definecolor{dorange}{rgb}{0.8,.4,.0}
\definecolor{green}{rgb}{0.0,1.0,0.0}
\definecolor{darkgreen}{rgb}{0.0,0.6,0.0}
\definecolor{purple}{rgb}{.4,.0,.4}
\definecolor{lightgrey}{rgb}{0.7, 0.7, 0.7}
\definecolor{grey}{rgb}{0.4, 0.4, 0.4}

\newcommand{\nc}{\newcommand}
\nc{\rnc}{\renewcommand} \nc{\ket}[1]{\left | \, #1 \right \rangle}
\nc{\bra}[1]{\left \langle #1 \, \right |}
\nc{\ua}{\uparrow} \nc{\da}{\downarrow}

\nc{\braket}[2]{\langle\, #1\,|\,#2\,\rangle}
\nc{\half}{\frac{1}{2}}

\nc{\prj}{\mathcal{P}} \nc{\hilb}{\mathcal{H}}
\nc{\pth}{\mathcal{C}} \nc{\inprod}[2]{\braket{#1}{#2}}
\nc{\upket}{\ket{\uparrow}} \nc{\downket}{\ket{\downarrow}}
\nc{\upbra}{\bra{\uparrow}} \nc{\downbra}{\bra{\downarrow}}

\begin{document}


\title{Surface growth scheme for bulk reconstruction and tensor network}

\author{Yi-Yu Lin$^1$} \email{linyy27@mail2.sysu.edu.cn}
\author{Jia-Rui Sun$^{1}$} \email{sunjiarui@mail.sysu.edu.cn}
\author{Yuan Sun$^{1}$} \email{sunyuan6@mail.sysu.edu.cn}

\affiliation{${}^1$School of Physics and Astronomy, Sun Yat-Sen University, Guangzhou 510275, China}


\begin{abstract}
We propose a surface growth approach to reconstruct the bulk spacetime geometry, motivated by Huygens' principle of wave propagation. We show that our formalism can be explicitly realized with the help of the surface/state correspondence and the one-shot entanglement distillation (OSED) method. We first construct a tensor network corresponding to a special surface growth picture with spherical symmetry and fractal feature using the OSED method and show that the resulting tensor network can be identified with the MERA-like tensor network, which gives a proof that the MERA-like tensor network is indeed a discretized version of the time slice of AdS spacetime, rather than just an analogy. Furthermore, we generalize the original OSED method to describe more general surface growth picture by using of the surface/state correspondence and the generalized RT formula, which leads to a more profound interpretation for the surface growth process and provides a concrete and intuitive way for the idea of entanglement wedge reconstruction.
\end{abstract}


\maketitle
\tableofcontents

\section{Introduction}
The anti-de Sitter/conformal field theory (AdS/CFT) correspondence and the general gauge/gravity duality not only provide powerful tools for studying strongly coupled quantum field theories, but also provide new perspectives for studying quantum gravity, in which the quantum field theory on the boundary is viewed as the quantum gravity description of the dual bulk gravity, instead of quantizing the gravity directly~\cite{Maldacena:1997re,Gubser:1998bc,Witten:1998qj}. To fully understand the duality between the bulk and the boundary, it is important to study how the bulk gravitational theory can be constructed from the boundary quantum field theory, this formalism is called the bulk reconstruction~\cite{Harlow:2018fse,Hamilton:2006az,Kabat:2017mun,Jafferis:2015del,Faulkner:2017vdd,Bao:2019bib,Faulkner:2018faa,Agon:2020mvu}. It has been shown that the notion of quantum entanglement, especially the holographic description of entanglement entropy, i.e. the holographic entanglement entropy, plays an essential role in the bulk reconstruction. It was proposed by Ryu and Takayanagi (RT) and later generalized by Hubeny, Rangamani and Takayanagi (HRT) that the entanglement entropy of a boundary subregion $A$ of a CFT is given by~\cite{Ryu:2006bv,Ryu:2006ef,Hubeny:2007xt}
\be\label{rt}{S_A} = \frac{{\rm Area}\left( {\gamma _A}\right)}{4G},\ee
where ${{\gamma _A}}$ is the codimension-two bulk minimal surface which anchored on the boundary of $A$ and homologous to $A$, which can be called the RT surface, and $G$ is the gravitational coupling constant. Clearly, the RT formula eq.(\ref{rt}) shows connections between the boundary entanglement entropy with the bulk geometry, which indicates that the entanglement structure of boundary state may play an important role in the emergence of the bulk theory.

In the project of bulk reconstruction, another important notion is the subregion duality~\cite{Czech:2012bh}, that is, given a boundary subregion $A$, one can use the CFT operators within $A$ rather than an entire boundary timeslice to reconstruct the bulk operators in a specific bulk subregion corresponding to $A$. It has been proposed that the appropriate bulk subregion dual to boundary subregion $A$ is the so-called entanglement wedge, which is defined as the bulk domain of dependence of any homology hypersurface (i.e., a spatial codimension-one submanifold bounded by ${{\gamma _A}}$ and $A$) for the bulk minimal surface ${{\gamma _A}}$. The entanglement wedge reconstruction was originally conjectured in~\cite{Headrick:2014cta,Czech:2012bh,Wall:2012uf}, and established in~\cite{Jafferis:2015del,Dong:2016eik,Cotler:2017erl} using the ideas of~\cite{Faulkner:2013ana}. However, the existing derivations of entanglement wedge reconstruction have been indirect, i.e., it relies on shedding the light of modern quantum information theory onto the holographic principle.

Here we propose a novel and interesting surface growth scheme to provide a direct and concrete way to reconstruct the bulk spacetime in the entanglement wedge. Our scheme is inspired by the fascinating picture of Huygens' principle of wave propagation and based on the surface/state correspondence. The surface/state correspondence suggests that the extremal surfaces anchored on any bulk regions have definite physical meanings similar to the RT surfaces anchored on the boundary of spacetime~\cite{Miyaji:2015yva,Miyaji:2015fia}. Our surface growth scheme assumes that new minimal surfaces can still grow into bulk from the minimal surfaces which anchored on the boundary, and so on, just as each sub-surface acting as a new source for the next minimal surface to anchor on. We will show that the boundary can detect and reconstruct the bulk information in this way, and it naturally leads us to a specific kind of corresponding tensor network, which in turn provides a reasonable and quantitative description for the surface growth scheme.

Tensor network is a $D+1$ dimensional discrete geometry, originally used in condensed matter physics as an effective approximation of $D$ dimensional quantum states. In particular, the multiscale entanglement renormalization ansatz (MERA) tensor network~\cite{Vidal:2007hda,Vidal:2008zz} is a tensor network representation of the ground state of critical quantum spin chains, which is extended by a dimension corresponding to the scale. In the AdS/CFT correspondence, the extra holographic dimension in AdS spacetime can be considered as renormalization group scale in the dual field theory. Therefore, from the holographic perspective, MERA was conjectured to describe a canonical time slice of AdS spacetime, i.e., the hyperbolic plane~\cite{Swingle:2009bg,Swingle:2012wq,Beny:2011vh,Czech:2015kbp}. Later on, it was argued that MERA on the real line should be interpreted as a Poincare patch of
de Sitter spacetime~\cite{SinaiKunkolienkar:2016lgg,Bao:2017qmt}. Recently, by using the path integral geometry~\cite{Milsted:2018vop,Milsted:2018yur}, \cite{Milsted:2018san} showed that the tensor network corresponding to the hyperbolic plane should be a variant of MERA tensor network, namely, the Euclidean MERA, while the traditional MERA describes neither the hyperbolic plane nor de Sitter spacetime, but an intermediate geometry, namely, a light sheet. In addition to the geometrical connections, the more important dynamical connections between the tensor network and the bulk gravity, namely, connections between the boundary Schr\"{o}dinger equation and the bulk Wheeler-DeWitt equation has also been revealed~\cite{Sun:2019ycv}. For more recent studies on relations between tensor network and holography, see e.g.~\cite{Pastawski:2015qua,Hayden:2016cfa,Qi:2013caa,Ling:2019akz,Ling:2018ajv,Ling:2018vza,Bhattacharyya:2017aly,
Bhattacharyya:2016hbx,Hung:2019zsk,Gan:2017nyt,Nomura:2018kji,Murdia:2020iac}.

In the present paper, we will use the method proposed in~\cite{Bao:2018pvs} to relate a tensor network with our surface growth scheme.\footnote{In this paper we will focus on the case of AdS$_3$.} The method mainly develops the idea of one-shot entanglement distillation (OSED), which states that the entanglement between subregions of some physical state can be distilled out of the state as a large number of EPR pairs, which can then be considered as the maximally entangled bonds of a projected entangled pair states (PEPS) network. We first construct a tensor network corresponding to a special surface growth picture with spherical symmetry and fractal feature, and will show that the resulting tensor network can be (approximately) identified with the MERA-like tensor network. As it has been proved in \cite{Bao:2018pvs}, the tensor network constructed by this method can reproduce the correct boundary state with high fidelity, and have a bulk geometry that matches the bulk AdS spacetime perfectly. Therefore, if our tensor network is (approximately) equivalent to MERA-like network, we provide a way to prove that the geometry of MERA-like network is indeed a discrete version of AdS spacetime, rather than just an analogy. Moreover, we further extend the original OSED method to describe more general surface growth picture in the framework of surface/state correspondence~\cite{Miyaji:2015yva}. It turns out that this generalization leads to a more profound understanding of the surface growth process and a meaningful conclusion, i.e., the grown extremal surfaces within the entanglement wedge of a boundary subregion can indeed detect the information inside the boundary subregion, which provides a concrete and intuitive way for the entanglement wedge reconstruction.

This paper is organized as follows. In section \ref{sec-intro}, we introduce and discuss the surface growth scheme qualitatively. In section \ref{sec-tn}, we construct a tensor network corresponding to a special surface growth scheme using OSED method. In section \ref{sec-tnl}, we argue that the tensor network constructed in the previous section can be identified with the MERA-like tensor network. In senction \ref{sec-sg}, we generalize the original OSED method in the framework of surface/state correspondence to adapt the more general surface growth scheme. Finally we give conclusions and discussions in section \ref{sect:conclusion}.


\section{Introduction to surface growth scheme}\label{sec-intro}
\begin{figure}[htbp]     \begin{center}
\includegraphics[height=7cm,clip]{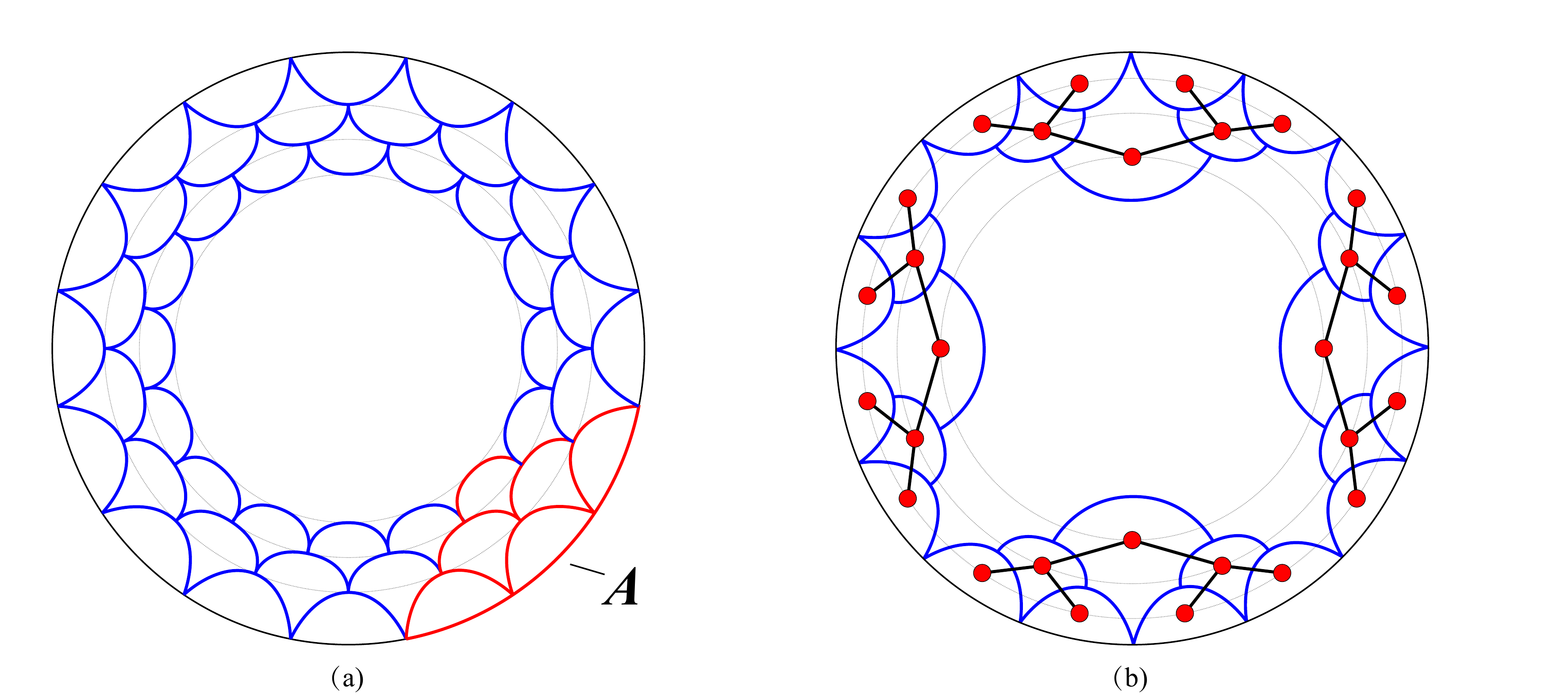}
\caption{(a) A surface growth picture for AdS$_3$, each two adjacent RT surfaces can generate a new minimal surface of the next layer, and so on. Note that a boundary subregion $A$ composed of $n$ pieces, can only grow $n-1$ new small surfaces in the second layer, and so on, therefore the entire $A$ region is expected to grow only as deep as $n$ layers in the bulk. (b) From another point of view, each component unit can be considered as a tensor. Furthermore, if we choose the surface growth picture appropriately, the resulting tensor network seems to correspond to the MERA-like tensor network.}
\label{fig1}
\end{center}	
\end{figure}

The basic idea of the surface growth proposal is as follows. In the present paper we will take the AdS$_3$ spacetime as an example,\footnote{For higher dimensional cases, although in principle the OSED procedure itself is independent of the spacetime dimensions, the generalization of OSED procedure to apply to the surfaces deep inside the bulk (as we will see later) does rely on the surface/state correspondence, which, however, is best supported in low dimensions, and less well supported in higher dimensions. We will leave the more detailed investigation about this topic in the future work.} its metric in the global coordinate is
\be\label{global} ds^2 = d\rho^2+{L^2}\left(- \cosh^2\frac{\rho}{L}dt^2 + \sinh^2\frac{\rho}{L}d\phi^2 \right),\ee
where $L$ is the curvature radius of the AdS spacetime and the AdS boundary is located at $\rho = \rho_0 \to \infty$ and its spatial section is just a circle. While in the Poincar\'{e} coordinate, the metric is
\be\label{Poincare} ds^2=L^2\left(\frac{dz^2-dt^2+dx^2}{z^2}\right),
\ee
where the boundary is located at $z=a\to 0$. Next, let us divide the spatial slice on the boundary into $N$ (which is large) equal segments in FIG.\ref{fig1}(a). According to RT formula, each small boundary subregion at least detects the nonlocal information about the global configuration of its corrsponding RT surface. Now, if we consider two adjacent RT surfaces, and take one point on each surface, such as its midpoint as in FIG.\ref{fig1}(a), then one can image that from these two adjacent points, a new minimal surface can continues to grow (or can be generated) so as to detect the information of deeper bulk region. By iterating this process, it seems that the initial boundary regions can probe the information deeper and deeper into the bulk, which means that the information in these bulk regions can be reconstructed. This picture is very similar to Huygens' principle of wave propagation, that is, every sub-surface becomes a new ``wave source'', just like the ripple or bubble growing.

\begin{figure}[htbp]     \begin{center}
		\includegraphics[height=10cm,clip]{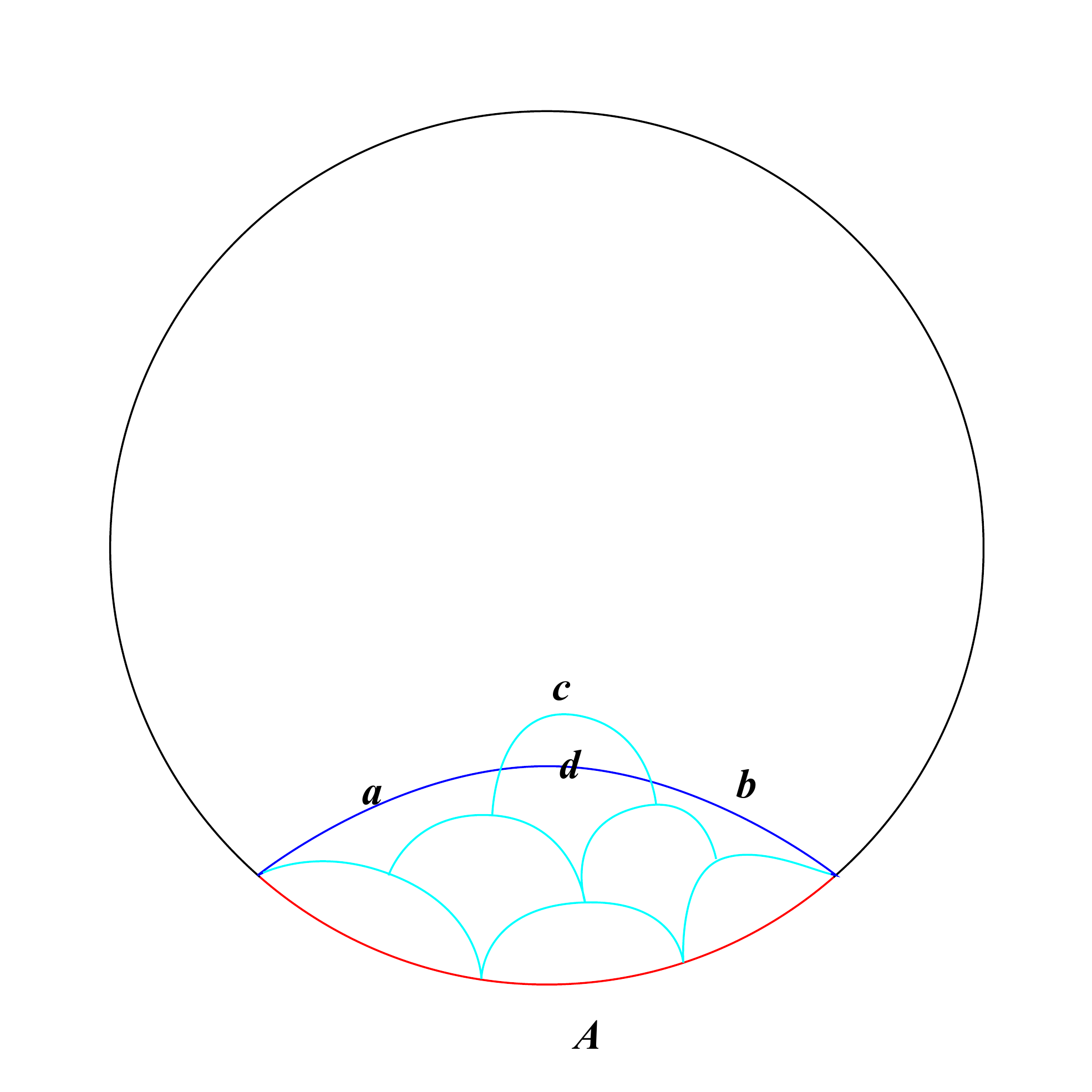}
		\caption{The bulk minimal surfaces growing out of a boundary subregion $A$ will not exceed the entanglement wedge of $A$
		}
		\label{contradiction}
	\end{center}	
\end{figure}

For this surface growth scheme with special symmetry in FIG.\ref{fig1}(a), intuitively, a boundary subregion $A$ composed of $n$ pieces can only grow $n$ small surfaces in the first layer, and $n-1$ new small surfaces in the second layer, and so on. By the time we get to the $n$th layer, there is only one small surface. Thus the entire $A$ region is expected to grow only as deep as $n$ layers in the bulk. Furthermore, it is easy to prove by contradiction that in the more general surface growth picture, all of the bulk minimal surfaces growing out of $A$ are also contained within the entanglement wedge of $A$. As shown in FIG.\ref{contradiction}, supposing (1), $adb$ is the RT surface corresponding to $A$, and (2), there exists a bulk minimal surface extending to the outside of the entanglement wedge of $A$, i.e., the outside of the RT surface of $A$, and we denote the part beyond the RT surface as $c$. Since $c$ is part of the bulk minimal surface, one should obtain $c<d$, and thus $acb<adb$, but this is contradictory to (1), according to which, the $adb$ surface should be the minimal surface corresponding to $A$. Therefore, the bulk minimal surfaces growing out of $A$ cannot exceed the entanglement wedge of $A$, rather, they should be contained within it. Moreover, if the endpoints for new growth are choosen to located closer to the boundaries of the previous surfaces, and $N$ is taken as a very large number, the envelope of these small minimal surfaces that grow out of $A$ should be exactly the minimal surface directly corresponding to $A$, because the perturbation of each small segment of minimal surfaces in the figure will make the length of entire envelople larger. Therefore, on the one hand, the bulk minimal surfaces growing out of a boundary subregion $A$ will not exceed the entanglement wedge of $A$, and on the other hand, it can reach anywhere in the entanglement wedge. It can thus be considered that the information of these bulk regions can be detected and reconstructed from the initial boundary regions in this way. Moreover, as long as the extension of the minimal surface of concern is small enough, this detection is approximately ``local'' (at least in the classical sense). In this paper, we will show that this surface growth scheme of bulk reconstruction can be described reasonably and quantitatively, and thus provide a concrete and intuitive way for the entanglement wedge reconstruction.

We can demonstrate the rationality of this idea qualitatively at least for the regions near the spacetime boundary as follows: since the minimal surface just extends a small distance in the radial direction away from the boundary, it can be considered that what we do is just taking the boundary cutoff a little larger and regarding the envelope of these minimal surfaces as a new boundary, thus we are just probing the bulk information at the new boundary. Actually, in the framework of surface/state correspondence \cite{Miyaji:2015yva,Miyaji:2015fia}, a generalized version of RT formula has been proposed, which suggests that the extremal surfaces in any bulk regions have definite physical meanings. More specifically, the surface/state correspondence claims that, if we consider an arbitrary closed convex surface $\Sigma$ in the bulk spacetime and a subregion ${\Sigma _A}$ of $\Sigma$, then this closed surface $\Sigma$ and the surface ${\Sigma _A}$ will correspond to quantum states described by density matrices $\rho \left( \Sigma  \right)$ and $\rho \left( {{\Sigma _A}} \right)$ respectively, and the entanglement entropy $S_A^\Sigma $ of subregion ${\Sigma _A}$ with respect to the quantum state $\rho \left( \Sigma  \right)$, i.e., the von Neumann entropy of $\rho \left( {{\Sigma _A}} \right)$ can be calculated by the area of the extremal surface $\gamma _A^\Sigma $ anchored on the boundary of ${\Sigma _A}$, i.e.
\be\label{gene0} S_A^\Sigma =\frac{{\rm Area}\left( {\gamma _A^\Sigma } \right)}{4G}.
\ee

Actually, it was shown by recent studies that these bulk extremal surfaces can also be investigated equivalently from the viewpoint of cutoff-AdS/$T\bar T$-deformed CFT correspondence (cAdS/dCFT)~\cite{Chen:2019mis,McGough:2016lol}. By considering a family of theories ${\mathcal{T}^{\left( \mu  \right)}}$ obtained by deforming a 2d CFT using the $T\bar T$ operator \cite{Zamolodchikov:2004ce}, cAdS/dCFT duality claims that the deformed theory  ${\mathcal{T}^{\left( \mu  \right)}}$ is dual to a gravitational theory living in a finite region in AdS$_3$ with a corresponding radial cutoff ${r_c}\left( \mu  \right)$, where $\mu$ denotes the deformation parameter. In other words, as the boundary moves into the bulk, one obtains a series of codimension-one hypersurfaces which are exactly where the $T\bar T$-deformed CFTs live and their corresponding codimension-two time slices associated with the quantum states of the $T\bar T$-deformed CFTs. It was thus proposed that the surface/state correspondence can be realized by the cAdS/dCFT correspondence to some extent, as long as one regards the states correponding these co-dimension two time slices in the framework of surface/state correspondence as vacuum states of the $T\bar T$-deformed CFTs in the framework of cAdS/dCFT duality~\cite{Chen:2019mis}. In addition, it has been also argued that the generalized RT formula still holds in cAdS/dCFT correspondence, in which the area of the extremal surface ending on the cut-off surface measures the holographic entanglement entropy in the $T\bar T$-deformed CFT~\cite{Chen:2018eqk,Donnelly:2018bef,Geng:2019ruz,Gorbenko:2018oov,Murdia:2019fax}. And this is exactly consistent with our qualitative arguement.

However, we would like to point out that the scheme proposed in this paper is not exactly the same as the above two pictures. The objects we study directly are extremal surfaces in the bulk spacetime, which grow directly on the extremal surfaces of the previous layer, and so on, while it is not necessary to introduce the auxiliary closed surfaces as in surface/state correspondence or cAdS/dCFT correspondence. We can intuitively name our picture as surface growth scheme. More specifically, as will be shown below, we will use the so-called OSED method~\cite{Bao:2018pvs} to implement our surface growth scheme.

\section{Constructing the tensor network}\label{sec-tn}
So far we have proposed a “surface growth” scheme to reconstruct the bulk spacetime. However, from another point of view, each component unit, namely, each sub-surface, can be considered as a tensor. Furthermore, intuitively, if we choose the surface growth picture appropriately, as shown in FIG.{\ref{fig1}}(b), then we can obtain a corresponding tensor network which seems to be equivalent to the traditional MERA tensor network.

Actually, a general procedure called one-shot entanglement distillation (OSED) has been proposed in~\cite{Bao:2018pvs} to construct a tensor network, which can approximate the boundary state with high accuracy, while the underlying geometry of the network can match the discrete version of dual bulk spacetime. In a sense, its construction method is also somewhat a picture of “growth”. Intuitively, it constructs a complete tensor network by adding the tensors inductively according to some certain order of a serious of nonintersecting RT surfaces, which cut the bulk into many pieces. Each piece bounded by the RT surfaces in the bulk corresponds to a tensor in the final network, whose legs corresponds to the RT surfaces bound the piece. Notice that all of these bulk extremal surfaces here grow out directly from the boundary of spacetime, and this is equivalent to grow out from the particular points very close to the ends of the previous extremal surfaces. The principle of OSED construction is briefly reviewed as follows.

In quantum information theory, for a state ${\left| \varphi  \right\rangle ^{ \otimes m}}$  obtained by the direct product of a large $m$ number of copies of an arbitrary quantum state $\left| \varphi  \right\rangle  \in {\mathcal{H}_A} \otimes {\mathcal{H}_{{A^c}}}$, there exists an operation called $entanglement ~distillation$, which can approximate it with high fidelity as a state described by a large $n$ number of Bell pairs manifestly encoding the entanglement between $A$ and its complement ${A^c}$, where $n$ satisfies a fixed asymptotic ratio relation $\frac{n}{m} \approx \frac{{S\left( A \right)}}{{\ln 2}}$ and $S(A)$ denotes the entanglement entropy of $A$, i.e.~\cite{hayden0204,Bao:2018pvs}
\be\label{simi}{\left| \varphi  \right\rangle ^{ \otimes m}} \approx \left( {W \otimes V} \right)\left[ {\frac{1}{{\sqrt D }}\sum\limits_{i = 0}^{{e^{S\left( A \right)m - O\left( {\sqrt m } \right)}}} {\left| {i\bar i} \right\rangle }  \otimes \sum\limits_{j = 0}^{{e^{O\left( {\sqrt m } \right)}}} {\sqrt {{p_j}} \left| {j\bar j} \right\rangle } } \right],\ee
where $W$ and $V$ are isometries that embedded these auxiliary Hilbert spaces of size ${{e^{S\left( A \right)m - O\left( {\sqrt m } \right)}}}$ and ${{e^{O\left( {\sqrt m } \right)}}}$ along with their complex conjugate Hilbert spaces, back into the physical space ${\mathcal{H}_A} \otimes {\mathcal{H}_{{A^c}}}$. In particular, since it is widely believed that having a gravity dual that looks like Einstein gravity coupled to matter requires a CFT with strong coupling and a large number of degrees of freedom (``large $N$''), \cite{Bao:2018pvs} demonstrated that this kind of semiclassical holographic limit of a single holographic state $\left| \Psi  \right\rangle $ can play the same information-theoretic role as the limit of a large number of identical copies of a single, non-holographic state $\left| \varphi  \right\rangle $ in non-holographic quantum information theory. More explicitly, it was shown that due to the holographic limit, for a reduced CFT density matrix ${\rho _A}$ of a holographic CFT full and pure state $\left| \Psi  \right\rangle $ describing a subregion $A$ of the CFT, one can always find a so-called ``smoothed state'' $\rho _A^\varepsilon$ to approximate it with very high fidelity, which is obtained by performing a ``smoothing'' operation on the original state ${\rho _A}$, and satisfying
\be\begin{array}{l}
{\rm rank}\left( {\rho _A^\varepsilon } \right) = {e^{{S_A} + O\left( {\sqrt {{S_A}} } \right)}},\\
{p _{\max }}\left( {\rho _A^\varepsilon } \right) = {e^{ - {S_A} + O\left( {\sqrt {{S_A}} } \right)}},
\end{array}\ee
where ${p _{\max }}\left( {\rho _A^\varepsilon } \right)$ is the largest eigenvalue of $\rho _A^\varepsilon$, and ${S_A}$ denotes the entanglement entropy of $A$, i.e., the von Neumann entropy of ${\rho _A}$. Note that the original subregion state ${\rho _A}$ can be described by the Schimidt decomposition of the original full pure state $\left| \Psi  \right\rangle $,
\be\left| \Psi  \right\rangle  = \sum\limits_n {\sqrt {{p _n}} {{\left| n \right\rangle }_A}} {\left| n \right\rangle _B},\ee
where $\left\{ {{p _n}} \right\}$ are the eigenvalues of the reduced density matrix ${\rho _A}$ of subregion $A$  and that of its complement ${\rho _B}$ simultaneously. Then the first key point of OSED is to describe the smoothed state $\rho _A^\varepsilon$ by the similar decomposition of the corresponding smoothed full state $\left| {{\Psi ^\varepsilon }} \right\rangle$ which also approximates the original full state $\left| \Psi  \right\rangle $ with high accuracy,
\be\left| {{\Psi ^\varepsilon }} \right\rangle  = \sum\limits_n {\sqrt {{{\tilde p }_n}} {{\left| n \right\rangle }_A}} {\left| n \right\rangle _B,}\ee
with $\left\{ {{{\tilde p }_n}} \right\}$ are the
eigenvalues of the smoothed state $\rho _A^\varepsilon $. Next, one can rearrange these basis eigenstates such that their probability spectrum ${{{\tilde p }_n}}$ is monotonically decreasing, i.e. ${{\tilde p }_{n + 1}} \le {{\tilde p }_n}$
, and break the resulting sum into blocks of size $\Delta $, which will be determined as $\Delta  = {e^{{S_A} - O(\sqrt {{S_A}} )}}$ later. One can thus take the average of the eigenvalues of each block as ${{{\tilde p }^{{\rm avg}}}_{n\Delta }}$, and approximate the smoothed state as~\cite{Bao:2018pvs}
\be\label{formula1}
	\left| {{\Psi ^\varepsilon }} \right\rangle  &=& \sum\limits_{n = 0}^{{\rm rank}[\rho _A^\varepsilon ]/\Delta  - 1} {\sum\limits_{m = 0}^{\Delta  - 1} {\sqrt {{{\tilde p }_{n\Delta  + m}}} {{\left| {n\Delta  + m} \right\rangle }_A}{{\left| {n\Delta  + m} \right\rangle }_B}} } \nno\\
	&=& \sum\limits_{n = 0}^{{e^{O(\sqrt S )}}} {\sum\limits_{m = 0}^{{e^{S - O(\sqrt S )}}} {\sqrt {{{\tilde p }^{{\rm avg}}}_{n\Delta }} {{\left| {n\Delta  + m} \right\rangle }_A}{{\left| {n\Delta  + m} \right\rangle }_B}} }
.\ee

\begin{figure}[htbp]     \begin{center}
\includegraphics[height=10cm,clip]{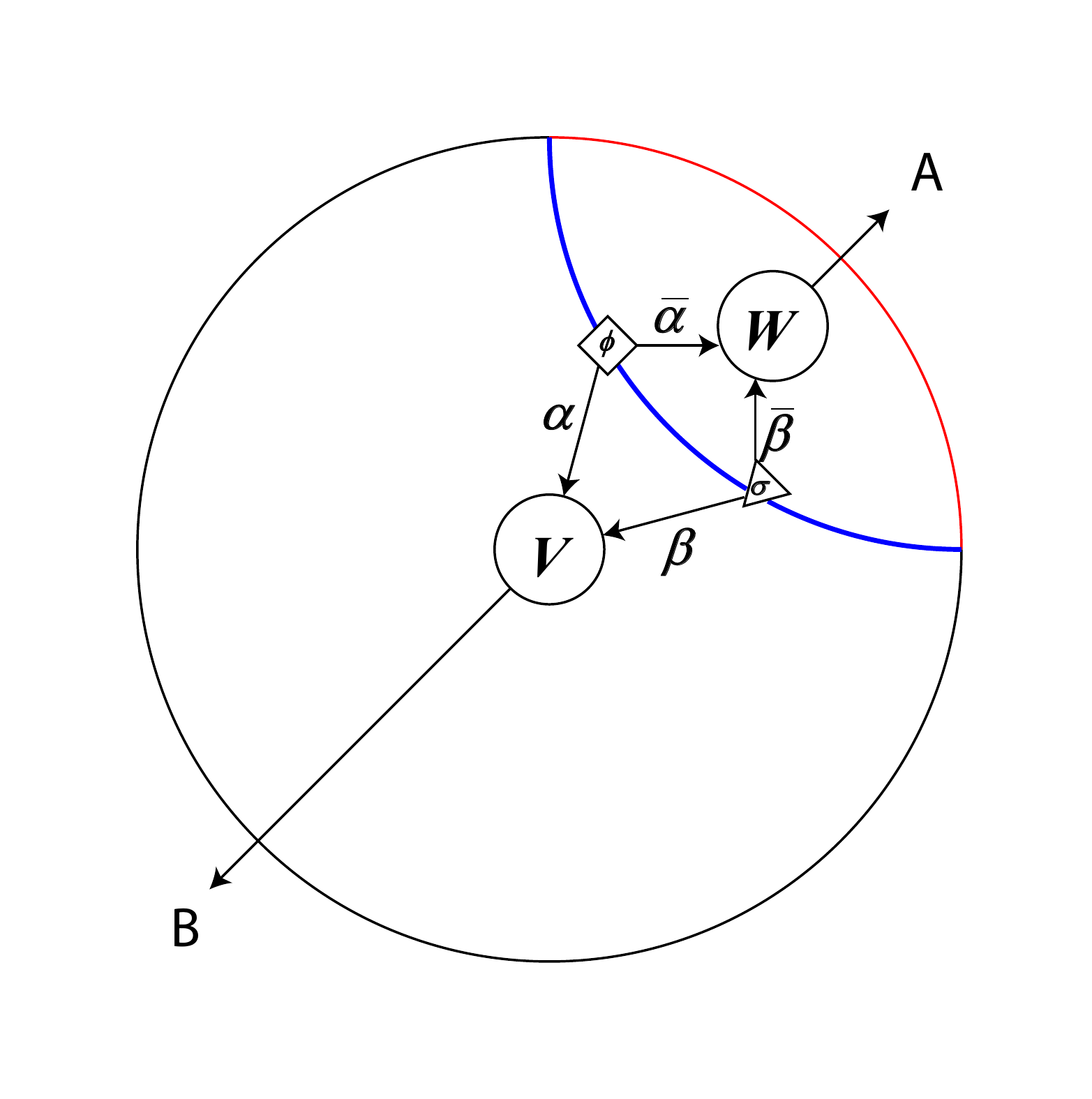}\caption{ The graphical representation of the OSED procedure for a holographic bipartite state.  By iterating this procedure on a holographic state inductively, one can construct a tensor network for the state. }
\label{fig2a}
\end{center}	
\end{figure}

Finally, one obtains the following tensor representation of the full holographic state, which is quite similar to the eq.(\ref{simi}). (From now on, we will omit the $\varepsilon $ for convenience, and adopt a convention that up-indices denote outward-pointing legs of tensor, while down-indices denote inward-pointing legs.)
\be{\Psi ^{AB}} = V_{\beta\alpha }^BW_{\bar \beta\bar \alpha }^A{\phi ^{\alpha \bar \alpha }}{\sigma ^{\beta\bar \beta}} = (V \otimes W)(\left| \phi  \right\rangle \otimes \left| \sigma  \right\rangle) \ee
where a maximally entangled state $\left| \phi  \right\rangle \otimes \left| \sigma  \right\rangle$ has been constructed (``distilled out''), with
\be\label{def1}\begin{array}{l}
	\left| \phi  \right\rangle  = \sum\limits_{m = 0}^{{e^{S - O(\sqrt S )}}} {{{\left| {m\bar m} \right\rangle }_{\alpha \bar \alpha }}}, \\
	\left| \sigma  \right\rangle  = \sum\limits_{n = 0}^{{e^{O(\sqrt S )}}} {\sqrt {\tilde p _{n\Delta }^{avg}} {{\left| {n\bar n} \right\rangle }_{\beta\bar \beta}}}.
\end{array}\ee
and their Hilbert space dimensions are
\be\label{def2}\begin{array}{l}
	{{\mathop{\rm dim \mathcal{H}}\nolimits} _\alpha } = {e^{S - O(\sqrt S )}},\\
	{{\mathop{\rm dim \mathcal{H}}\nolimits} _\beta} = {e^{O(\sqrt S )}}.
\end{array}\ee
And the isometris $V$ and $W$, which correspond to the entanglement wedge of $B$ and $A$ respectively, map the arbitrarily chosen bases of these auxiliary Hilbert spaces and the corresponding bases in their complex conjugate Hilbert spaces into the eigenstates of the smoothed states ${\rho _B}$ and ${\rho _A}$ respectively, namely,
\be\begin{array}{l}
	V:{\mathcal{H}_\beta} \otimes {\mathcal{H}_\alpha } \to {\mathcal{H}_B},\\
	V{\left| n \right\rangle _\beta}{\left| m \right\rangle _\alpha } = {\left| {n\Delta  + m} \right\rangle _B}.
\end{array}\ee
and
\be\begin{array}{l}
	W:{\mathcal{H}_{\bar \beta}} \otimes {\mathcal{H}_{\bar \alpha }} \to {\mathcal{H}_A},\\
	W{\left| n \right\rangle _{\bar \beta}}{\left| m \right\rangle _{\bar \alpha }} = {\left| {n\Delta  + m} \right\rangle _A}.
\end{array}\ee
Obviously, according to eq.(\ref{def2}), the logarithm of the bond dimension, i.e. the Hilbert space dimension of $\left| \phi  \right\rangle $ corresponds to the entanglement entropy of $A$ exactly, while $\left| \sigma  \right\rangle $ should be interpreted as the quantum fluctuation. The graphical representation of this procedure is shown in FIG.\ref{fig2a}. We thus accomplish the one-shot entanglement distillation (OSED) for a holographic state, and \cite{Bao:2018pvs} shows that by iterating the OSED procedure on a holographic state, we can construct a PEPS-style tensor network for the state. If one considers this holographic state as the boundary state of some dual bulk spacetime, then the underlying geometry of the tensor network matches with the geometry of the bulk spacetime~\cite{Bao:2018pvs}.

\begin{figure}[htbp]     \begin{center}
		\includegraphics[height=10cm,clip]{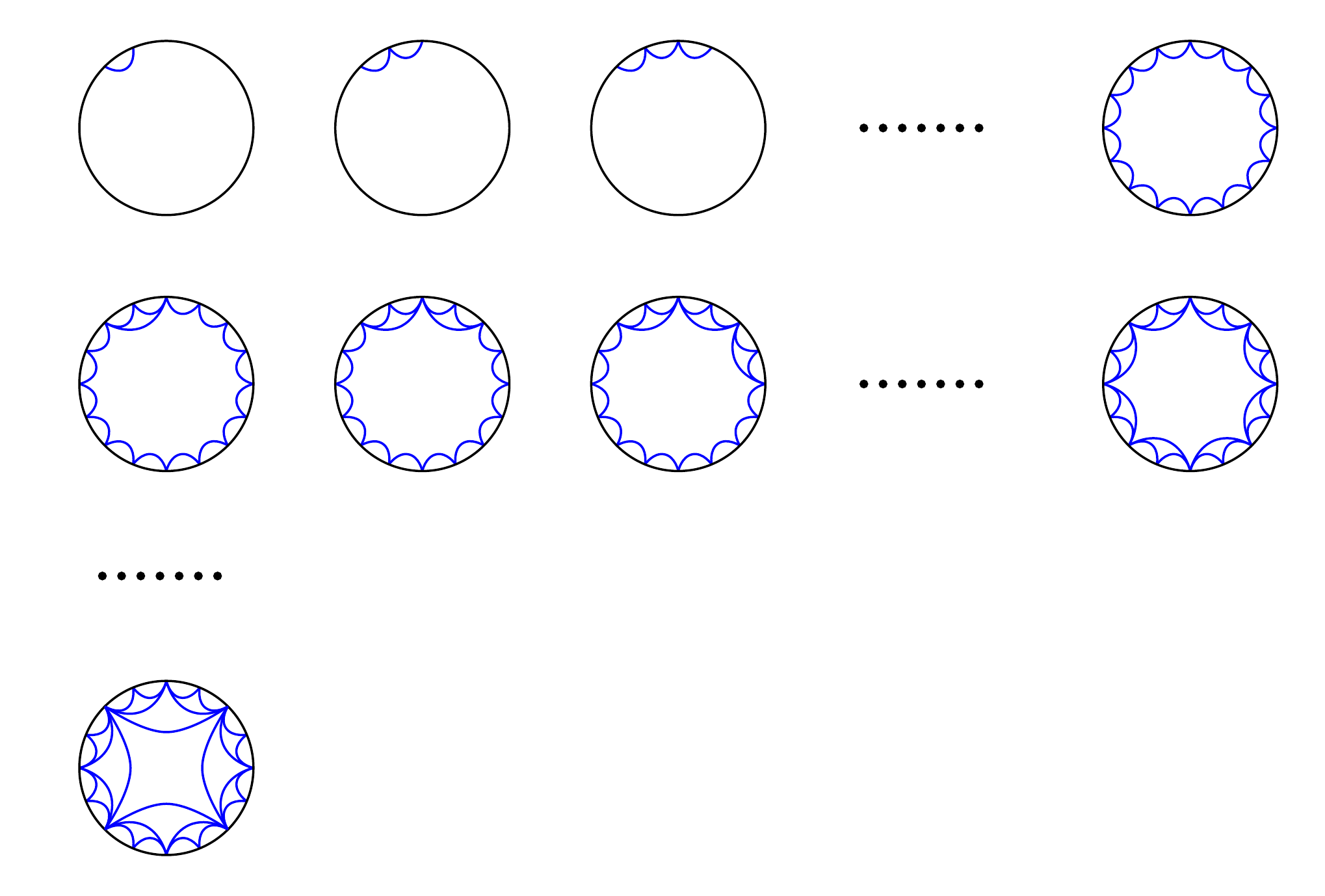}\caption{We propose a scheme to construct a corresponding tensor network for a pure AdS$_3$ spacetime. We use $N$ identical RT surfaces to discretize the first layer step by step. Then, we discretize the second layer similarly, where each minimal surface of the second layer bounds two adjacent small RT surfaces in the first layer exactly. Then we discretize the third layer in the same way, and so on. }
		\label{fig2b}
	\end{center}	
\end{figure}

Now, using this OSED method , we propose a scheme as shown in FIG.\ref{fig2b} to construct a corresponding tensor network for the bulk spacetime. We will consider the case when the bulk spacetime is pure AdS$_3$, which is holographically dual to a 2d CFT vacuum state on the boundary. Firstly, we use $N$ identical RT surfaces to discretize the first layer step by step. Then, we discretize the second layer similarly, where each RT surface of the second layer bounds two adjacent small RT surfaces in the first layer exactly. Then we discretize the third layer in the same way, and so on. Suppose that the first layer has ${\rm{N}} = {2^n}$ petals of RT surfaces, after numerous steps of segmentation, finally we can obtain a satisfactory discretization as shown in the figure.

\begin{figure}[htbp]     \begin{center}
\includegraphics[height=8cm,clip]{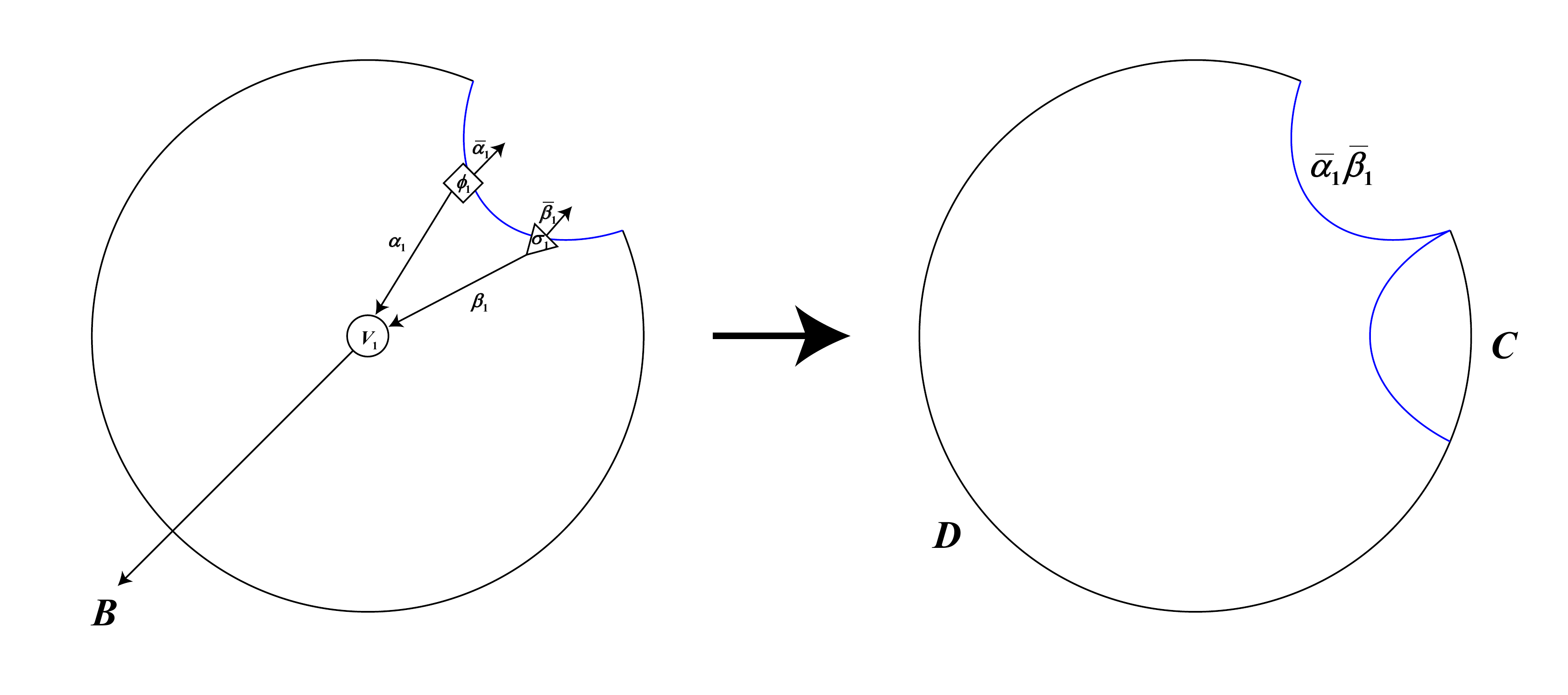}
\caption{By defining a subregion state ${\Psi _V}$ of the remaining region that has been cut off a small piece, we can apply the same bipartite distillation procedure to ${\Psi _V}$. }
\label{fig3a}
\end{center}	
\end{figure}

\begin{figure}[htbp]     \begin{center}
		\includegraphics[height=8cm,clip]{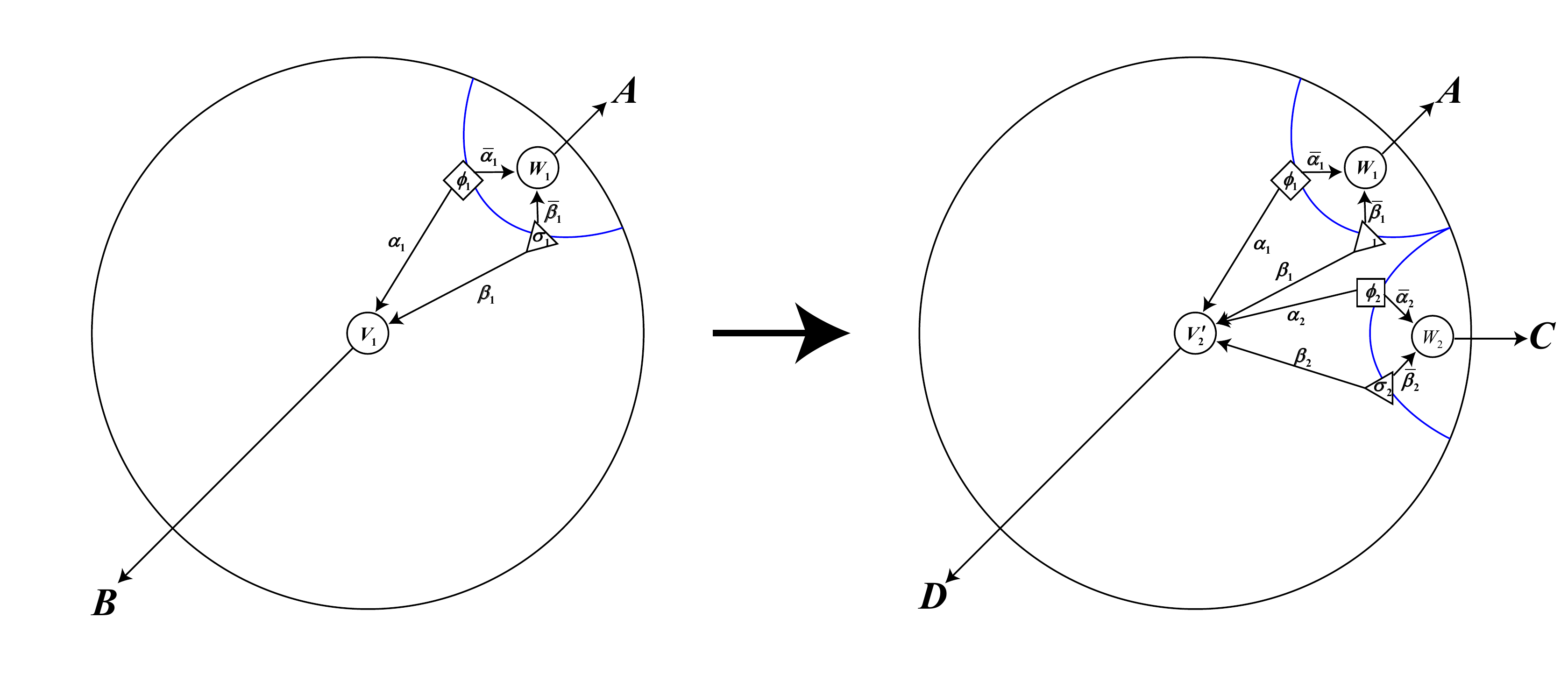}
		\caption{The resulting tensor network is obtained by replacing the original ${V_1}$ tensor with ${V_1}_{{f_1}{\alpha _1}}^{B} = V_2^{'} {W_2}{\left| \phi  \right\rangle _2}{\left| \sigma  \right\rangle _2}$
		}
		\label{fig3b}
	\end{center}	
\end{figure}

Since the first step of the discretization has been carried out directly as shown in the FIG.\ref{fig2a}, we start by the second step of descretization. As shown in FIG.\ref{fig3a}, by defining a subregion state ${\Psi _V}$ for
the remaining region that has been cut off a small piece, we can apply the same bipartite distillation procedure to the subregion state. Defining
\be\begin{array}{l}
{\Psi ^{AB}} = V_{\beta\alpha }^BW_{\bar \beta\bar \alpha }^A{\phi ^{\alpha \bar \alpha }}{\sigma ^{\beta\bar \beta}}= {\Psi _V}^{B\bar \beta\bar \alpha }W_{\bar \beta\bar \alpha }^A,
\end{array}\ee
that is
\be\label{equ1}{\Psi _V}^{B\bar \beta\bar \alpha } = V_{\beta\alpha }^B{\phi ^{\alpha \bar \alpha }}{\sigma ^{\beta\bar \beta}} = V\left| \phi  \right\rangle \left| \sigma  \right\rangle. \ee
As shown in the FIG.\ref{fig3a}, we further decompose $B$ region into $C$ and $D$, thus ${\Psi _V}^{B\bar \beta\bar \alpha }$ can be decomposed in the same way as before
\be{\Psi _V}^{({{\bar \beta }_1}{{\bar \alpha }_1}D)C} = \sum\limits_{n = 0}^{{e^{O(\sqrt S )}}} {\sum\limits_{m = 0}^{{e^{S - O(\sqrt S )}}} {\sqrt {{{\tilde p }^{{\rm avg}}}_{n\Delta }} {{\left| {n\Delta  + m} \right\rangle }_C}{{\left| {n\Delta  + m} \right\rangle }_{{{\bar \beta }_1}{{\bar \alpha }_1}D}}}}. \ee
Similarly, we have
\be\label{equ2}{\Psi _V}^{({{\bar \beta }_1}{{\bar \alpha }_1}D)C} = {V_2}_{{\beta_2}{\alpha _2}}^{{{\bar \beta }_1}{{\bar \alpha }_1}D}{W_2}_{{{\bar \beta}_2}{{\bar \alpha }_2}}^C{\phi _2}^{{\alpha _2}{{\bar \alpha }_2}}{\sigma _2}^{{\beta_2}{{\bar \beta}_2}}.\ee
Combining eq.(\ref{equ1}) with eq.(\ref{equ2}), we have
\be{V_1}_{{\beta_1}{\alpha _1}}^B{\phi _1}^{{\alpha _1}{{\bar \alpha }_1}}{\sigma _1}^{{\beta_1}{{\bar \beta}_1}} = {V_2}_{{\beta_2}{\alpha _2}}^{{{\bar \beta }_1}{{\bar \alpha }_1}D}{W_2}_{{{\bar \beta}_2}{{\bar \alpha }_2}}^C{\phi _2}^{{\alpha _2}{{\bar \alpha }_2}}{\sigma _2}^{{\beta_2}{{\bar \beta}_2}},\ee
where we have added subscripts to distinguish the tensors associated to different steps. Note that there
is a canonical isomorphism between the states $\left| \phi  \right\rangle  \in {H_\alpha } \otimes {H_\alpha }$ and $\left| \sigma  \right\rangle  \in {H_\beta} \otimes {H_\beta}$ and
operators $\phi :{H_\alpha } \to {H_\alpha }$ and $\sigma :{H_\beta} \to {H_\beta}$. Since $\left| \phi  \right\rangle $ and $\left| \sigma  \right\rangle $ are full-rank, these operators are invertible. In other words, one can move ${\phi _1}^{{\alpha _1}{{\bar \alpha }_1}}{\sigma _1}^{{\beta_1}{{\bar \beta}_1}}$ to the right, and define
\be
{V_2}^{'} = {\phi _1}^{ - 1}{\sigma _1}^{ - 1}{V_2},\ee
that is
\be V_{2{\alpha _1}{\beta_1}{\alpha _2}{\beta_2}}^{'D} = {\left( {\phi _1^{ - 1}} \right)_{^{{\alpha _1}{{\bar \alpha }_1}}}}{\left( {\sigma _1^{ - 1}} \right)_{^{{\beta_1}{{\bar \beta}_1}}}}V_{2{\beta_2}{\alpha _2}}^{{{\bar \beta}_1}{{\bar \alpha }_1}D},\ee
thus, we have
\be{V_1}_{{\beta_1}{\alpha _1}}^{B} = V_2^{'} {W_2}{\left| \phi  \right\rangle _2}{\left| \sigma  \right\rangle _2}.\ee
Therefore, we find that after each step of discretization, we only need to replace the original $V$ tensor to obtain a new tensor network, just as shown in FIG.\ref{fig3b}, and we have
\be\left| \Psi  \right\rangle  = {V_1}{W_1}{\left| \phi  \right\rangle _1}{\left| \sigma  \right\rangle _1} = V_2^{'}{W_2}{\left| \phi  \right\rangle _2}{\left| \sigma  \right\rangle _2}{W_1}{\left| \phi  \right\rangle _1}{\left| \sigma  \right\rangle _1}.\ee

So far, it is easy to generalize this procedure to the subsequent steps. However, it should be noted that if one wants to construct the first layer of the exact MERA tensor network, one needs to ensure that the $W$ tensors associated with the boundary subregions of the same size are identical. Here we provide a rigorous proof, using the condition that $W$ is isometry, i.e. ${W^\dag }W = 1$. The proof involves writing the state as a matrix which corresponds to the tensor form. Explicitly, each probability amplitude ${c_{ij}}$ of the bipartite basis state is regarded as the matrix element. For convenience, we rewrite the full state as
\be\left| \Psi  \right\rangle  = \sum\limits_n {{c_{{a_n}{b_n}}}\left| {{a_n}} \right\rangle \left| {{b_n}} \right\rangle },
\ee
where ${\left| {{a_n}} \right\rangle }$ and ${\left| {{b_n}} \right\rangle }$ denote ${{{\left| n \right\rangle }_A}}$ and ${{{\left| n \right\rangle }_B}}$ respectively, and ${{c_{{a_n}{b_n}}}}$ denote ${\sqrt {{p _n}} }$, thus
\be\Psi  = \left( {\begin{array}{*{20}{c}}
	{{c_{{a_1}{b_1}}}}& \ldots &0\\
	\vdots & \ddots & \vdots \\
	0& \cdots &{{c_{{a_n}{b_n}}}}
	\end{array}} \right).\ee
In particular, $\Psi $ is a diagonal matrix. As for ${\Psi _V}$, we have
\be\left| {{\Psi _V}} \right\rangle  = \sum\limits_{n,i} {{c_{{b_n}i}}\left| {{b_n}i} \right\rangle }, \ee
in which $i = 1,2, \cdots ,s$ denotes the state at the ``notch'' after cutting off the small piece $A$, which is mapped to the ${\left| {{a_n}} \right\rangle }$ state of $A$ region boundary state by $W$ tensor. In the matrix form, we have
\be{\Psi _V} = \left( {\begin{array}{*{20}{c}}
	{{c_{{b_1}1}}}& \ldots &{{c_{{b_n}1}}}\\
	\vdots & \ddots & \vdots \\
	{{c_{{b_1}s}}}& \cdots &{{c_{{b_n}s}}}
	\end{array}} \right),\ee
and
\be
\Psi  &=& W{\Psi _V}\nno\\
&=& \left( {\begin{array}{*{20}{c}}
	{{w_{{a_1}1}}}& \ldots &{{w_{{a_1}s}}}\\
	\vdots & \ddots & \vdots \\
	{{w_{{a_n}1}}}& \cdots &{{w_{{a_n}s}}}
	\end{array}} \right)\left( {\begin{array}{*{20}{c}}
	{{c_{{b_1}1}}}& \ldots &{{c_{{b_n}1}}}\\
	\vdots & \ddots & \vdots \\
	{{c_{{b_1}s}}}& \cdots &{{c_{{b_n}s}}}
	\end{array}} \right).
\ee

Now, we want to prove that for an operator acting on $C$ alone, which is part of region $B$,
\be\left\langle {{\Psi _V}} \right|{O_C} \otimes 1\left| {{\Psi _V}} \right\rangle  = \left\langle \Psi  \right|{O_C} \otimes 1\left| \Psi  \right\rangle. \ee
Because
\be\left\langle {{\Psi _V}} \right|{O_C} \otimes 1\left| {{\Psi _V}} \right\rangle
= \sum\limits_s\left( {{{\left| {{c_{{b_1}s}}} \right|}^2}} \left\langle {{b_1}} \right|{O_C}\left| {{b_1}} \right\rangle  +  {{{\left| {{c_{{b_2}s}}} \right|}^2}} \left\langle {{b_2}} \right|{O_C}\left| {{b_2}} \right\rangle  +  \cdots  +  {{{\left| {{c_{{b_n}s}}} \right|}^2}} \left\langle {{b_n}} \right|{O_C}\left| {{b_n}} \right\rangle\right), \ee
and
\be\left\langle \Psi  \right|{O_C} \otimes 1\left| \Psi  \right\rangle  = {\left| {{c_{{a_1}{b_1}}}} \right|^2}\left\langle {{b_1}} \right|{O_C}\left| {{b_1}} \right\rangle  + {\left| {{c_{{a_2}{b_2}}}} \right|^2}\left\langle {{b_2}} \right|{O_C}\left| {{b_2}} \right\rangle  +  \cdots {\left| {{c_{{a_n}{b_n}}}} \right|^2}\left\langle {{b_n}} \right|{O_C}\left| {{b_n}} \right\rangle. \ee
Therefore, we just need to prove that
\be\sum\limits_s {{{\left| {{c_{{b_n}s}}} \right|}^2}}  = {\left| {{c_{{a_n}{b_n}}}} \right|^2}.\ee
Note that in the matrix form, this is just
\be{\Psi _V}^\dag {\Psi _V} = \left( {\begin{array}{*{20}{c}}
	{{{\left| {{c_{{a_1}{b_1}}}} \right|}^2}}& \ldots &0\\
	\vdots & \ddots & \vdots \\
	0& \cdots &{{{\left| {{c_{{a_n}{b_n}}}} \right|}^2}}
	\end{array}} \right) \equiv {\Psi ^\dag }\Psi. \ee
From ${W^\dag }W = 1$, one can immediately prove this by
\be{\Psi ^\dag }\Psi  = {\Psi _V}^\dag {W^\dag }W{\Psi _V} = {\Psi _V}^\dag {\Psi _V}.\ee
Similary, we can prove that
\be\left\langle {{\Psi _V}} \right|{O_B} \otimes 1\left| {{\Psi _V}} \right\rangle  = \left\langle \Psi  \right|{O_B} \otimes 1\left| \Psi  \right\rangle. \ee
Therefore, when we divide $B$ region into $C$ region and $D$ region in the pure state $\left| {{\Psi _V}} \right\rangle $, we can guarantee that the operation is unaffected by the previous cut by $W$ tensor. Thus, as long as the second boundary subregion is of the the same size as the first, the ${W_1}$ tensor and ${W_2}$ tensor are identical, so are ${\left| \phi  \right\rangle _1}{\left| \sigma  \right\rangle _1}$ and ${\left| \phi  \right\rangle _2}{\left| \sigma  \right\rangle _2}$. The same reasoning can be used in each of the following step, so we can safely proceed with the following discretization.

\begin{figure}[htbp]     \begin{center}
		\includegraphics[height=10cm,clip]{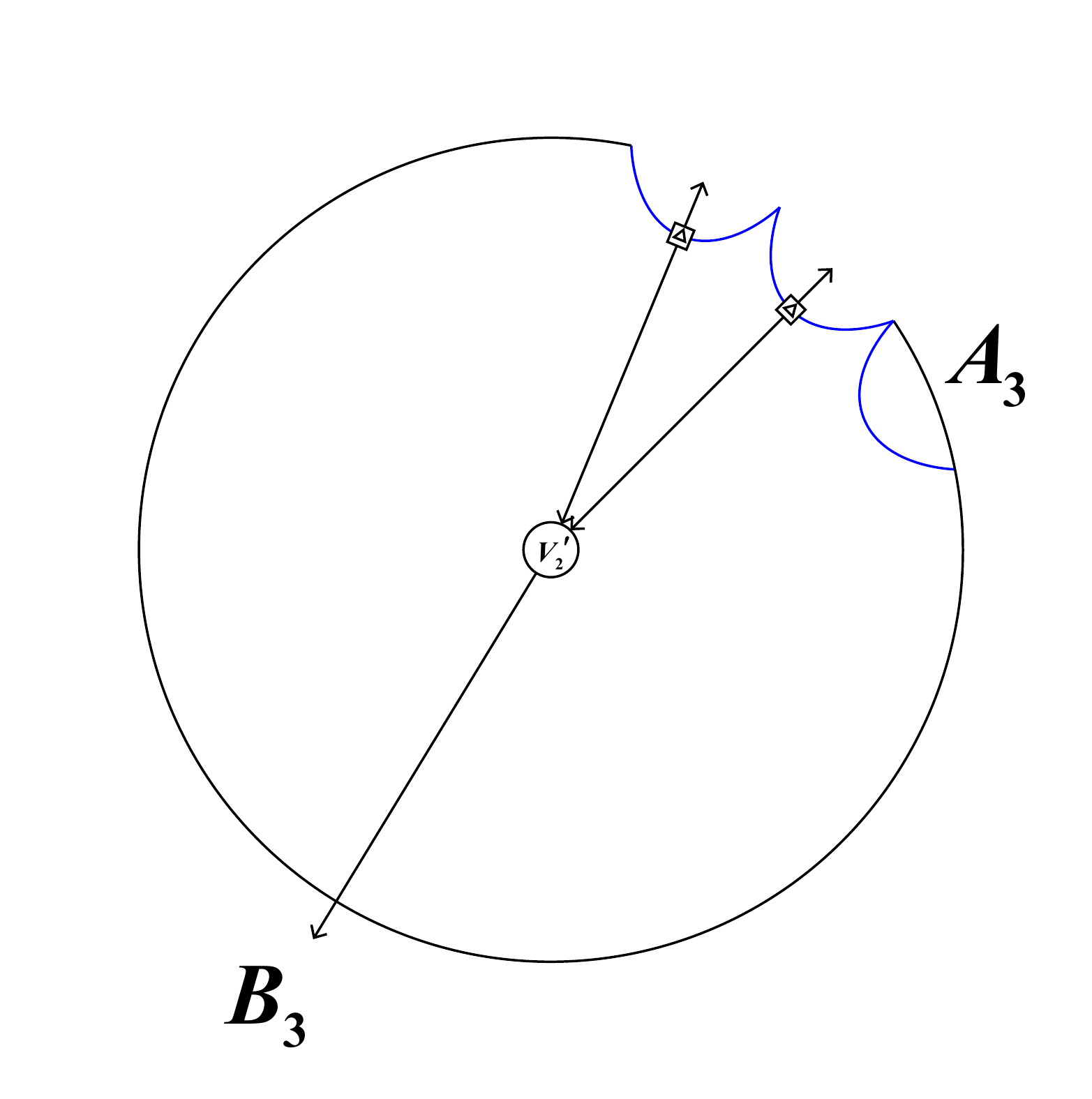}
		\caption{The third step of discretization.
		}
		\label{fig4a}
	\end{center}	
\end{figure}

Next, we move on to step three. Similarly, as shown in the FIG.\ref{fig4a}, we define the state associated to the remaining region after cutting off two identical small pieces as
\be\label{equ3}\Psi _{{V_2}}^{{B_3}{{\bar \beta}_1}{{\bar \alpha }_1}{{\bar \beta}_2}{{\bar \alpha }_2}{A_3}} = V_{2{\alpha _1}{\beta_1}{\alpha _2}{\beta_2}}^{'{B_2}}{\phi _1}^{{\alpha _1}{{\bar \alpha }_1}}{\sigma _1}^{{\beta_1}{{\bar \beta}_1}}{\phi _2}^{{\alpha _2}{{\bar \alpha }_2}}{\sigma _2}^{{\beta_2}{{\bar \beta}_2}},\ee
where for convenience, we will denoted $A$ in the previous section as ${A_1}$, and $C$ as ${A_2}$, and $D$ as ${B_2} \sim {B_3}{A_3}$. Accordingly, from the same logic, it can be decomposed as
\be\label{equ4}\Psi _{{V_2}}^{{B_3}{{\bar \beta}_1}{{\bar \alpha }_1}{{\bar \beta}_2}{{\bar \alpha }_2}{A_3}} = V_{3{\beta_3}{\alpha _3}}^{{B_3}{{\bar \beta}_1}{{\bar \alpha }_1}{{\bar \beta}_2}{{\bar \alpha }_2}}W_{3{{\bar \beta}_3}{{\bar \alpha }_3}}^{{A_3}}{\phi _3}^{{\alpha _3}{{\bar \alpha }_3}}{\sigma _3}^{{\beta_3}{{\bar \beta}_3}}.
\ee
Further combining eq.(\ref{equ3}) with eq.(\ref{equ4}), we can similarly define
\be V_{3{\alpha _1}{\beta_1}{\alpha _2}{\beta_2}}^{'{B_3}} = {\left( {\phi _1^{ - 1}} \right)_{^{{\alpha _1}{{\bar \alpha }_1}}}}{\left( {\sigma _1^{ - 1}} \right)_{^{{\beta_1}{{\bar \beta}_1}}}}{\left( {\phi _2^{ - 1}} \right)_{^{{\alpha _2}{{\bar \alpha }_2}}}}{\left( {\sigma _2^{ - 1}} \right)_{^{{\beta_2}{{\bar \beta}_2}}}}V_{3{\beta_3}{\alpha _3}}^{{B_3}{{\bar \beta}_1}{{\bar \alpha }_1}{{\bar \beta}_2}{{\bar \alpha }_2}}
\ee
and thus obtain
\be
V_{2{\alpha _1}{\beta_1}{\alpha _2}{\beta_2}}^{'{B_2}} = V_{3{\alpha _1}{\beta_1}{\alpha _2}{\beta_2}}^{'{B_3}}W_{3{{\bar \beta}_3}{{\bar \alpha }_3}}^{{A_3}}{\phi _3}^{{\alpha _3}{{\bar \alpha }_3}}{\sigma _3}^{{\beta_3}{{\bar \beta}_3}},
\ee
namely,
\be V_2^{'} = V_3^{'}{W_3}{\left| \phi  \right\rangle _3}{\left| \sigma  \right\rangle _3}.
\ee
Consequently, we can use $V_3^{'}{W_3}{\left| \phi  \right\rangle _3}{\left| \sigma  \right\rangle _3}$ to replace $V_2^{'}$ in the $\left| \Psi  \right\rangle $, and obtain
\be{\left| \Psi  \right\rangle _3} = V_3^{'}\prod\limits_{i = 1}^3 {{W_i}{{\left| \phi  \right\rangle }_i}{{\left| \sigma  \right\rangle }_i}}, \ee
where a subscript has been added to denote the step number.

\begin{figure}[htbp]     \begin{center}
		\includegraphics[height=10cm,clip]{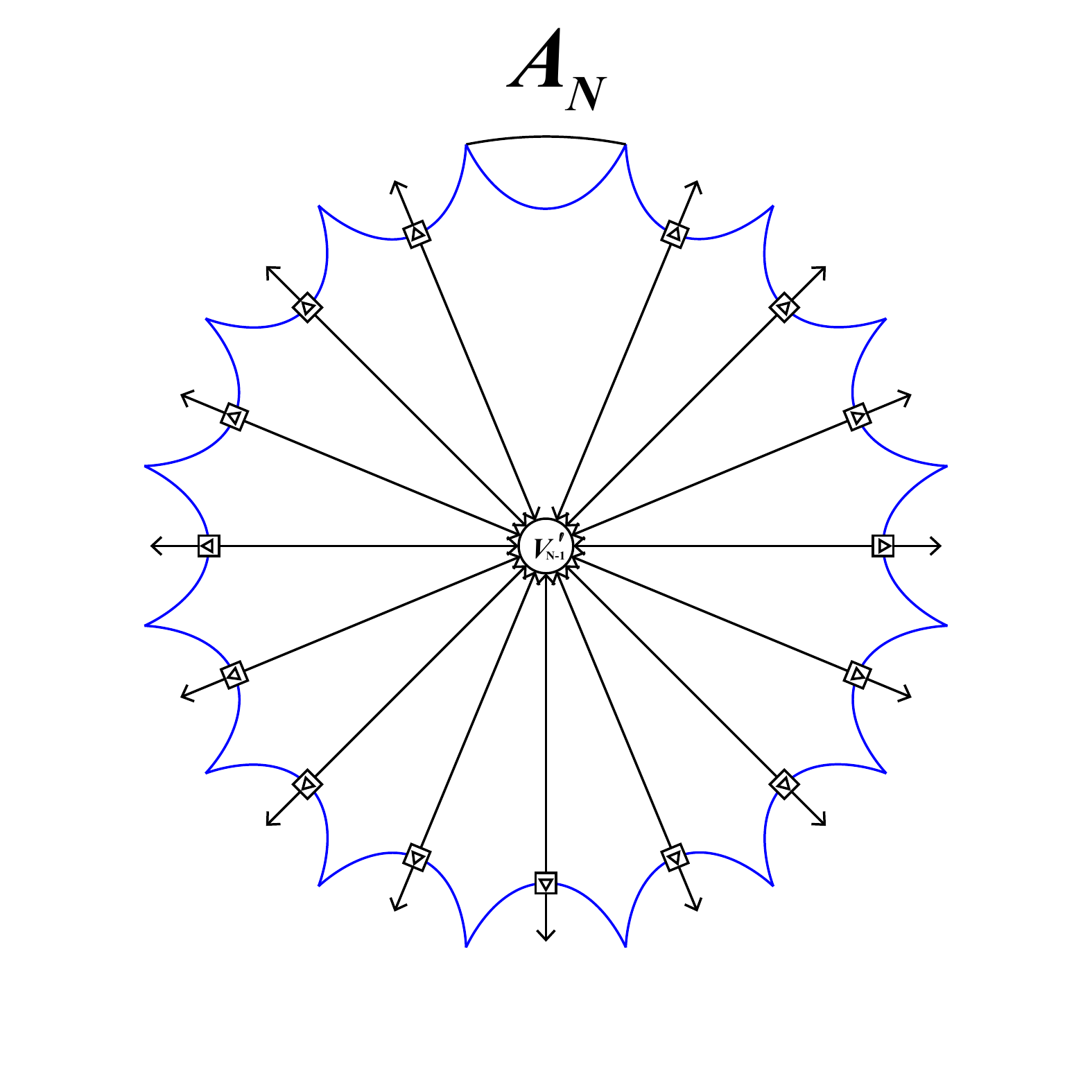}
		\caption{The last step of the first layer.
			}
		\label{fig4b}
	\end{center}	
\end{figure}

The third step can be immediately generalized to the last step of the first layer $N$. One just need to keep in mind that at each step the ${V'}$ tensor in the original tensor network is replaced by a new composite tensor, which includes a new ${V'}$ tensor, and brings a new $W$ tensor corresponding to this step of discretization. Explicitly, as shown in FIG.\ref{fig4b}, similarly, we define the state associated to the remaining region after cutting off $N - 1$ identical small pieces as
\be\label{equ5}{\Psi _{{V_{N - 1}}}} = V_{N - 1{ {\alpha _1}{\beta_1} \cdots {\alpha _{N - 1}}{\beta_{N - 1}}}
 }^{'{A_N}}\prod\limits_{i = 1}^{N - 1} {{\phi _i}^{{\alpha _i}{{\bar \alpha }_i}}{\sigma _i}^{{\beta_i}{{\bar \beta}_i}}}. \ee

Similarly, it can be decomposed as
\be\label{equ6}{\Psi _{{V_{N - 1}}}} = V_{N{\beta_N}{\alpha _N}}^{ {{{\bar \alpha }_1}{{\bar \beta}_1} \cdots {{\bar \alpha }_{N-1}}{{\bar \beta}_{N-1}}} }W_{N{{\bar \beta}_N}{{\bar \alpha }_N}}^{{A_N}}{\phi _N}^{{\alpha _N}{{\bar \alpha }_N}}{\sigma _N}^{{\beta_N}{{\bar \beta}_N}}.\ee
Combining eq.(\ref{equ5}) with eq.(\ref{equ6}), we can similarly define
\be V_{N{ {\alpha _1}{\beta_1} \cdots {\alpha _{N }}{\beta_{N }}} }^{'} = \prod\limits_{i = 1}^{N - 1} {{{\left( {\phi _i^{ - 1}} \right)}_{^{{\alpha _i}{{\bar \alpha }_i}}}}{{\left( {\sigma _i^{ - 1}} \right)}_{^{{\beta_i}{{\bar \beta}_i}}}}} V_{N{\beta_N}{\alpha _N}}^{ {{{\bar \alpha }_1}{{\bar \beta}_1} \cdots {{\bar \alpha }_{N-1}}{{\bar \beta}_{N-1}}} }\ee
and thus obtain
\be V_{N - 1}^{'} = V_N^{'}{W_N}{\left| \phi  \right\rangle _N}{\left| \sigma  \right\rangle _N}.\ee
Note that since at the final step of the first layer, ${B_{N - 1}} \sim {B_N}{A_N}$, while ${A_N} \sim {B_{N - 1}}$, thus ${B_N} \sim 0$, the $V_N^{'}$ tensor no longer has an up-index corresponding to the outward-pointing leg which directly map the ${{{\left| \phi  \right\rangle }_i}{{\left| \sigma  \right\rangle }_i}}$ states associated with the RT surfaces to the boundary state. Finally, using $V_N^{'}{W_N}{\left| \phi  \right\rangle _N}{\left| \sigma  \right\rangle _N}$ to replace $V_{N-1}^{'}$ in $\left| \Psi  \right\rangle $. Therefore, after the complete process of the first layer discretization, the full boundary state can be expressed as
\be\label{1st}{\left| \Psi  \right\rangle }^{{\rm 1st}} = V_N^{'{\rm 1st}}\prod\limits_{i = 1}^N {{W_i^{{\rm 1st}}}{{\left| \phi  \right\rangle }_i^{{\rm 1st}}}{{\left| \sigma  \right\rangle }_i^{{\rm 1st}}}}, \ee
in which ${\left| \phi  \right\rangle _i^{{\rm 1st}}\left| \sigma  \right\rangle _i^{{\rm 1st}}}$ represents the distilled state associated with the RT surface corresponds to subregion ${A_i}$, and ${W_i^{{\rm 1st}}}$ tensor maps the state represented by its $\bar \alpha \bar \beta$ leg to the reduced state on ${A_i}$, and the ${V_N^{'{\rm 1st}}}$ tensor connects all the remaining legs. Notice that we use the superscript 1st to denote the first layer.

\begin{figure}[htbp]     \begin{center}
		\includegraphics[height=8cm,clip]{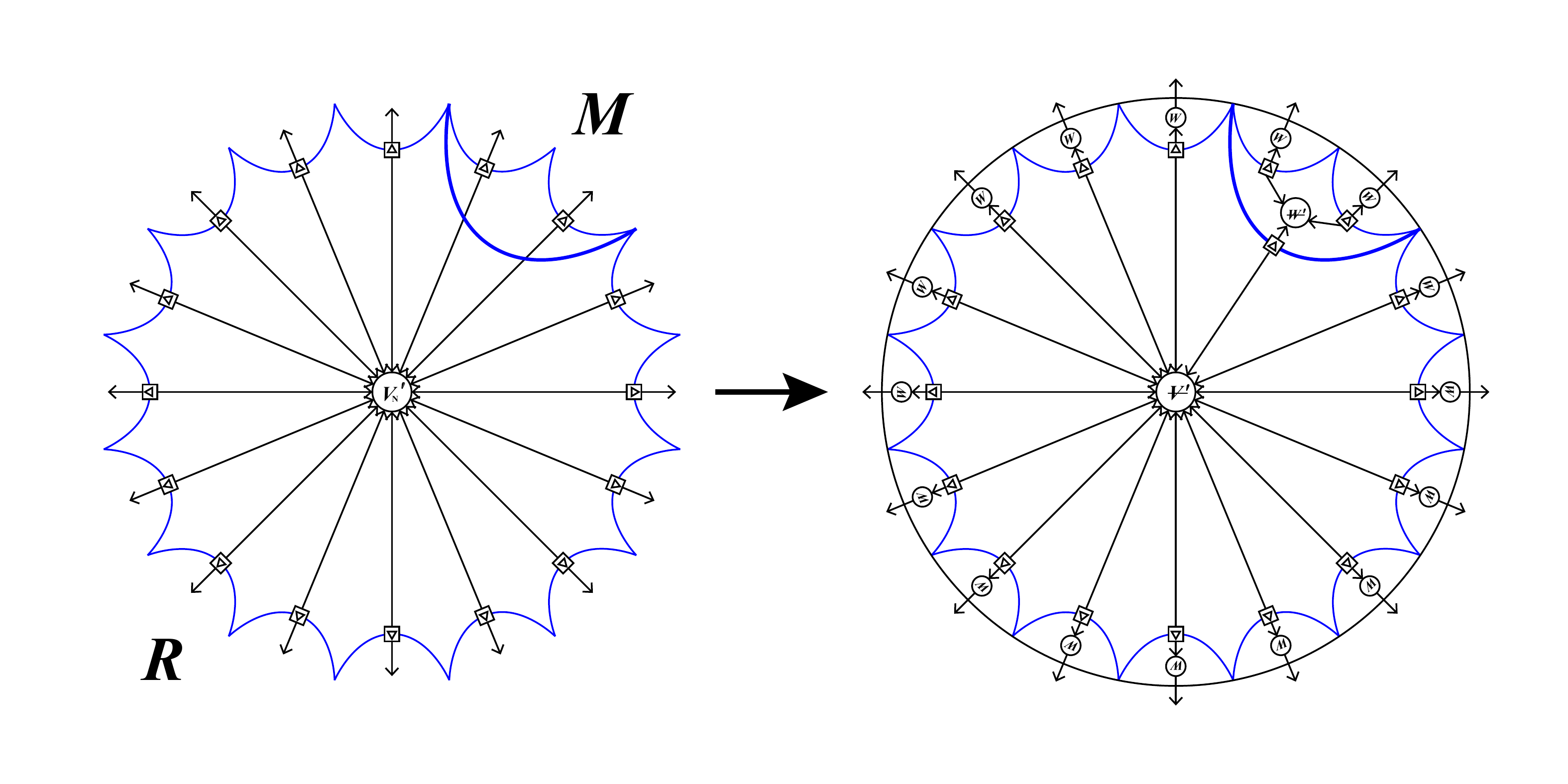}
		\caption{The first step of the second layer. Note that we have drawn the $\phi$ tensor and $\sigma$ tensor together for convenience.}
		\label{fig4c}
	\end{center}	
\end{figure}

The construction of the second layer is almost the same as the construction of the first layer in principle, except that the $W$ tensor needs to be redefined in a similar manner as $V$ tensor. We first define the state associated to the remaining region after cutting off all pieces of the first layer as
\be\label{equ7}{\Psi _{{V_N}}} = (V^{'{\rm 1st}}_N)_{ {\alpha _1}{\beta_1} \cdots  {\alpha _N}{\beta_N}}\prod\limits_{i = 1}^N {({\phi _i^{{\rm 1st}})}^{{\alpha _i}{{\bar \alpha }_i}}{(\sigma _i^{{\rm 1st}})}^{{\beta_i}{{\bar \beta}_i}}}. \ee
Similarly, dividing it by $M$ and its complement $R$, where $M \sim \prod\limits_{i = 1}^2 {{{\bar \alpha }_i}{{\bar \beta}_i}} $, $R \sim \prod\limits_{i = 3}^N {{{\bar \alpha }_i}{{\bar \beta}_i}} $, we obtain
\be\label{equ8}{\Psi _{{V_N}}} =  (V^{{\rm 2nd}})_{ \beta \alpha }^{R}  (W^{{\rm 2nd}})_{\bar \beta \bar \alpha }^{ M}({ \phi }^{{\rm 2nd}})^{ \alpha \bar \alpha }({ \sigma }^{{\rm 2nd}})^{ \beta \bar \beta},\ee
where we have used the superscript 2nd to denote the second layer. Combining eq.(\ref{equ7}) with eq.(\ref{equ8}) gives
\be{(V_N^{'{\rm 1st}})_{{\alpha _1}{\beta_1} \cdots {\alpha _N}{\beta_N}}} = \prod\limits_{i = 1}^N {(\phi _i^{{\rm 1st}})_{{\alpha _i}{{\bar \alpha }_i}}^{ - 1}(\sigma _i^{{\rm 1st}})_{{\beta_i}{{\bar \beta}_i}}^{ - 1}({V^{{\rm 2nd}}})_{\beta\alpha }^{{{\bar \alpha }_3}{{\bar \beta}_3} \cdots {{\bar \alpha }_N}{{\bar \beta}_N}}({W^{{\rm 2nd}}})_{\bar \beta\bar \alpha }^{{{\bar \alpha }_1}{{\bar \beta}_2}{{\bar \alpha }_2}{{\bar \beta}_2}}{{({\phi ^{{\rm 2nd}}})}^{\alpha \beta}}{{({\sigma ^{{\rm 2nd}}})}^{\beta\bar \beta}}}, \ee
and further redefining
\be{V^{'{\rm 2nd}}} = {({V^{'{\rm 2nd}}})_{\alpha \beta{\alpha _1}{\beta_1} \cdots {\alpha _3}{\beta_3}}} \equiv \prod\limits_{i = 3}^N {(\phi _i^{{\rm 1st}})_{{\alpha _i}{{\bar \alpha }_i}}^{ - 1}(\sigma _i^{{\rm 1st}})_{{\beta_i}{{\bar \beta}_i}}^{ - 1}({V^{{\rm 2nd}}})_{\alpha \beta}^{{{\bar \alpha }_3}{{\bar \beta}_3} \cdots {{\bar \alpha }_N}{{\bar \beta}_N}}}, \ee
and
\be\label{diff1}{W^{'{\rm 2nd}}} = {({W^{'{\rm 2nd}}})_{\bar \alpha \bar \beta{\alpha _1}{\beta_1} \cdots {\alpha _3}{\beta_3}}} \equiv \prod\limits_{i = 1}^2 {(\phi _i^{{\rm 1st}})_{{\alpha _i}{{\bar \alpha }_i}}^{ - 1}(\sigma _i^{{\rm 1st}})_{{\beta_i}{{\bar \beta}_i}}^{ - 1}({W^{{\rm 2nd}}})_{\bar \alpha \bar \beta}^{{{\bar \alpha }_1}{{\bar \beta}_1}{{\bar \alpha }_2}{{\bar \beta}_2}}}, \ee
one obtains
\be V_N^{'{\rm 1st}} = {V^{'{\rm 2nd}}}{W^{'{\rm 2nd}}}{\left| \phi  \right\rangle ^{{\rm 2nd}}}{\left| \sigma  \right\rangle ^{{\rm 2nd}}}.\ee
Its graphical representation is shown in FIG.\ref{fig4c}. The full boundary state then can be further expressed as
\be{\left| \Psi  \right\rangle } = {V^{'{\rm 2nd}}}{W^{'{\rm 2nd}}}{\left| \phi  \right\rangle ^{{\rm 2nd}}}{\left| \sigma  \right\rangle ^{{\rm 2nd}}}\prod\limits_{i = 1}^N {W_i^{{\rm 1st}}\left| \phi  \right\rangle _j^{{\rm 1st}}\left| \sigma  \right\rangle _j^{{\rm 1st}}}. \ee

The step of the second layer can be directly generalized to any step of any layer because of its apparent universality. We can always obtain the ${W^{'}}$ tensor for each layer, which always connects the ${\left| \phi  \right\rangle \left| \sigma  \right\rangle }$ states associated with the two adjacent minimal surfaces with the ${\left| \phi  \right\rangle \left| \sigma  \right\rangle }$ state associated with the minimal surface of the next layer. Actually, we will show that the ${W^{'}}$ tensor plays the role of both the disentangler as well as the coarse-grainer in the MERA tensor network in the next section.

\begin{figure}[htbp]     \begin{center}
		\includegraphics[height=10cm,clip]{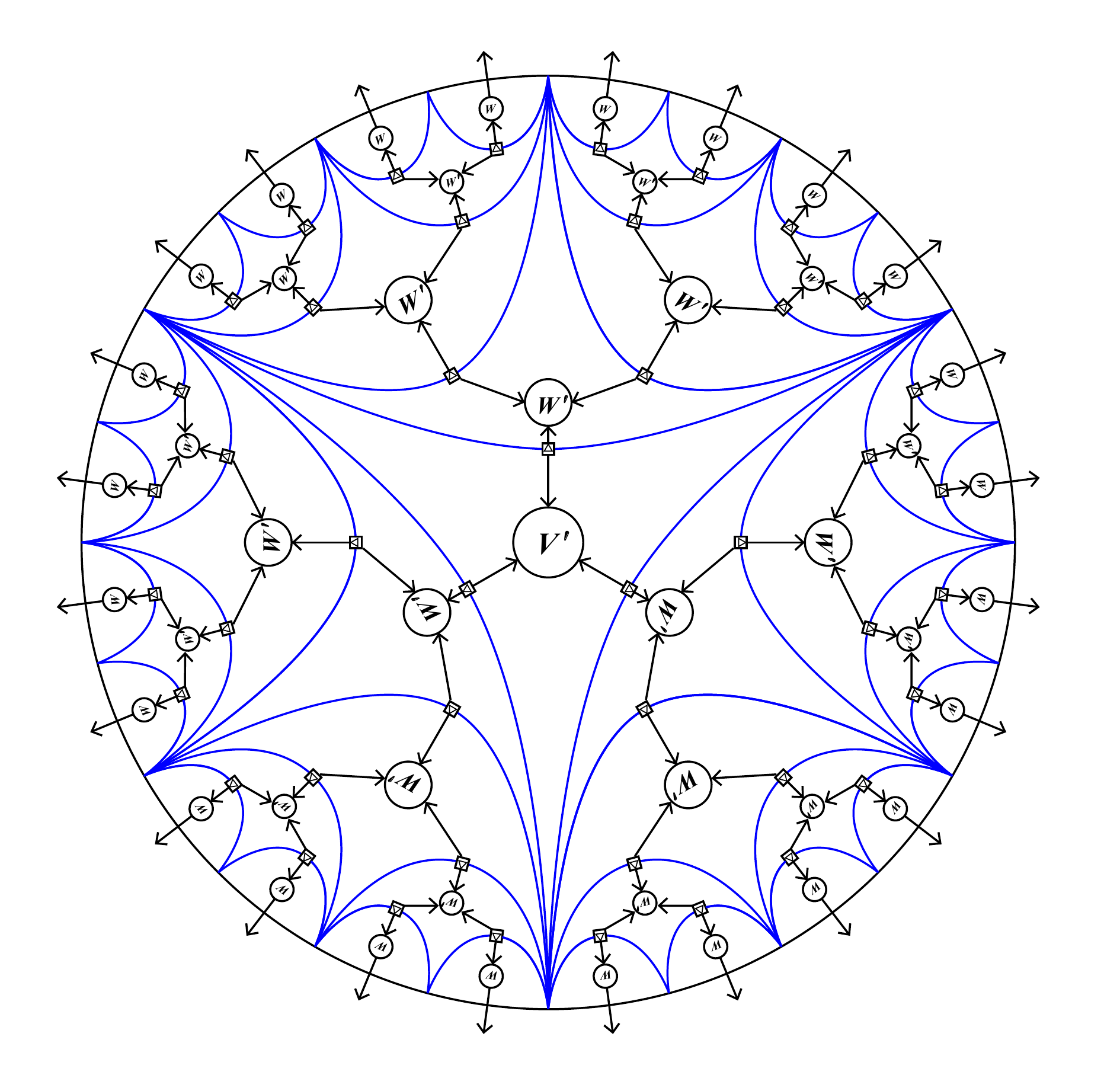}
		\caption{The resulting tensor network corresponding to our surface growth scheme.
			}
		\label{fig5a}
	\end{center}	
\end{figure}

According to this, after the full second layer of ``surface growth'', we obtain
\be{\left| \Psi  \right\rangle ^{{\rm 2nd}}} = V_M^{'{\rm 2nd}}(\prod\limits_{j = 1}^{M = N/2} {W_j^{'{\rm 2nd}}\left| \phi  \right\rangle _j^{{\rm 2nd}}\left| \sigma  \right\rangle _j^{{\rm 2nd}}})( \prod\limits_{i = 1}^N {W_i^{{\rm 1st}}\left| \phi  \right\rangle _j^{{\rm 1st}}\left| \sigma  \right\rangle _j^{{\rm 1st}}}), \ee
similarly, after the $k$-th layer of ``surface growth'', we obtain
\be{\left| \Psi  \right\rangle ^{k{\rm th}}} = V_M^{'k{\rm th}}\prod\limits_{i = 1}^{N/{2^{k - 1}}} {\prod\limits_{a = 1}^k {W_i^{'a{\rm th}}\left| \phi  \right\rangle _i^{a{\rm th}}\left| \sigma  \right\rangle _i^{a{\rm th}}} } \ee
Finally, we obtain the resulting tensor network as shown in FIG.\ref{fig5a}.

\begin{figure}[htbp]     \begin{center}
		\includegraphics[height=10cm,clip]{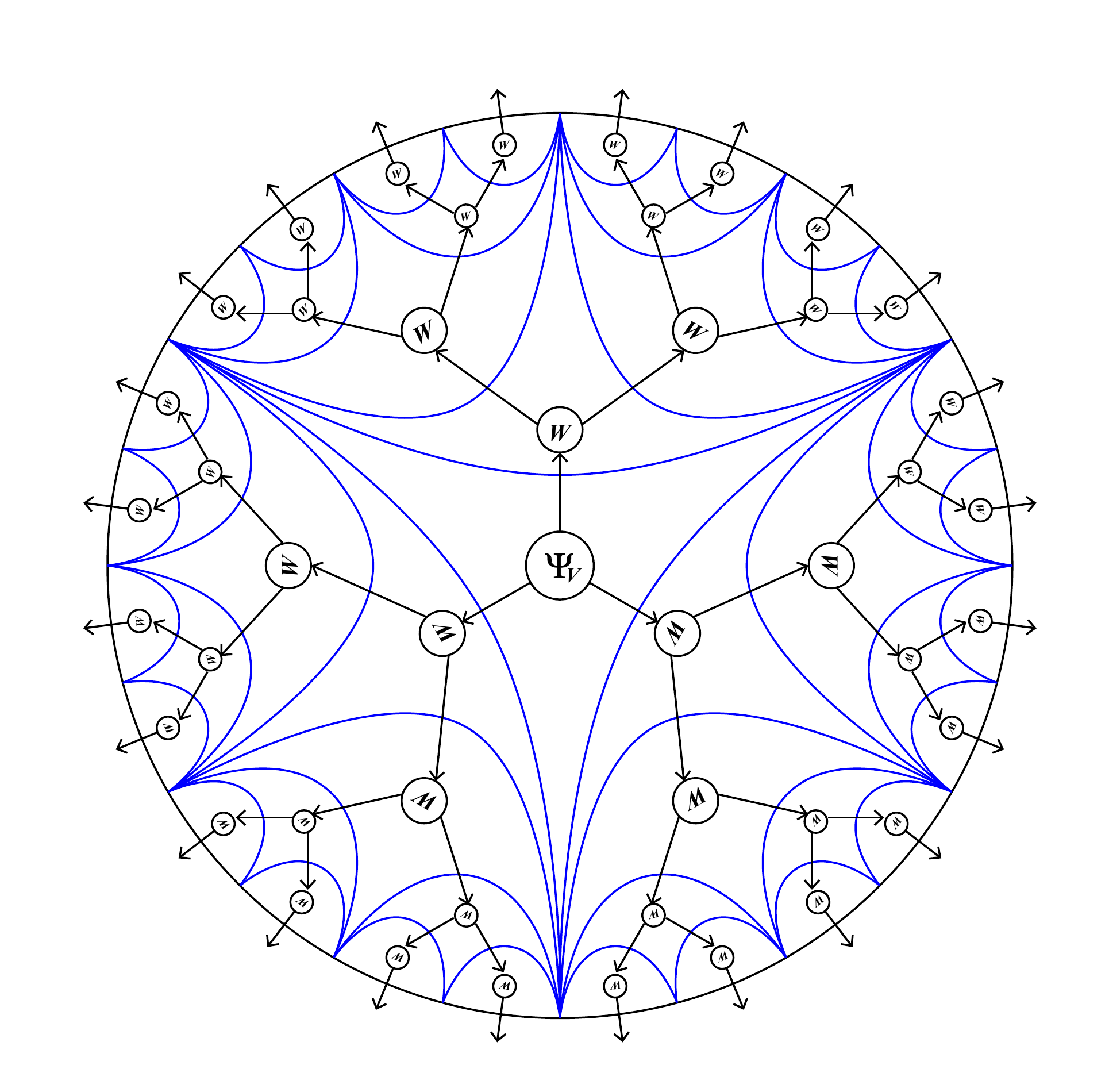}
		\caption{Another representation of the resulting tensor network, which is more appropriate to be identified with MERA-like tensor network.
			Note the difference of the component tensors of the two representations.
		}
		\label{fig5b}
	\end{center}	
\end{figure}

A comment: in this section, in the above representation of our tensor network, we have chosen to present the ${\left| \phi  \right\rangle \left| \sigma  \right\rangle }$ state tensor associated with each surface, since their dimensions clearly match the entanglement entropy. Nevertheless, in the next section, in order to accurately identify this network with the MERA-like tensor network, we will absorb these states into the ${V^{'}}$ tensors, as one can see in FIG.\ref{fig5b}.

\section{Identifying with MERA-like tensor network}\label{sec-tnl}
\begin{figure}[htbp]     \begin{center}
		\includegraphics[height=7.5cm,clip]{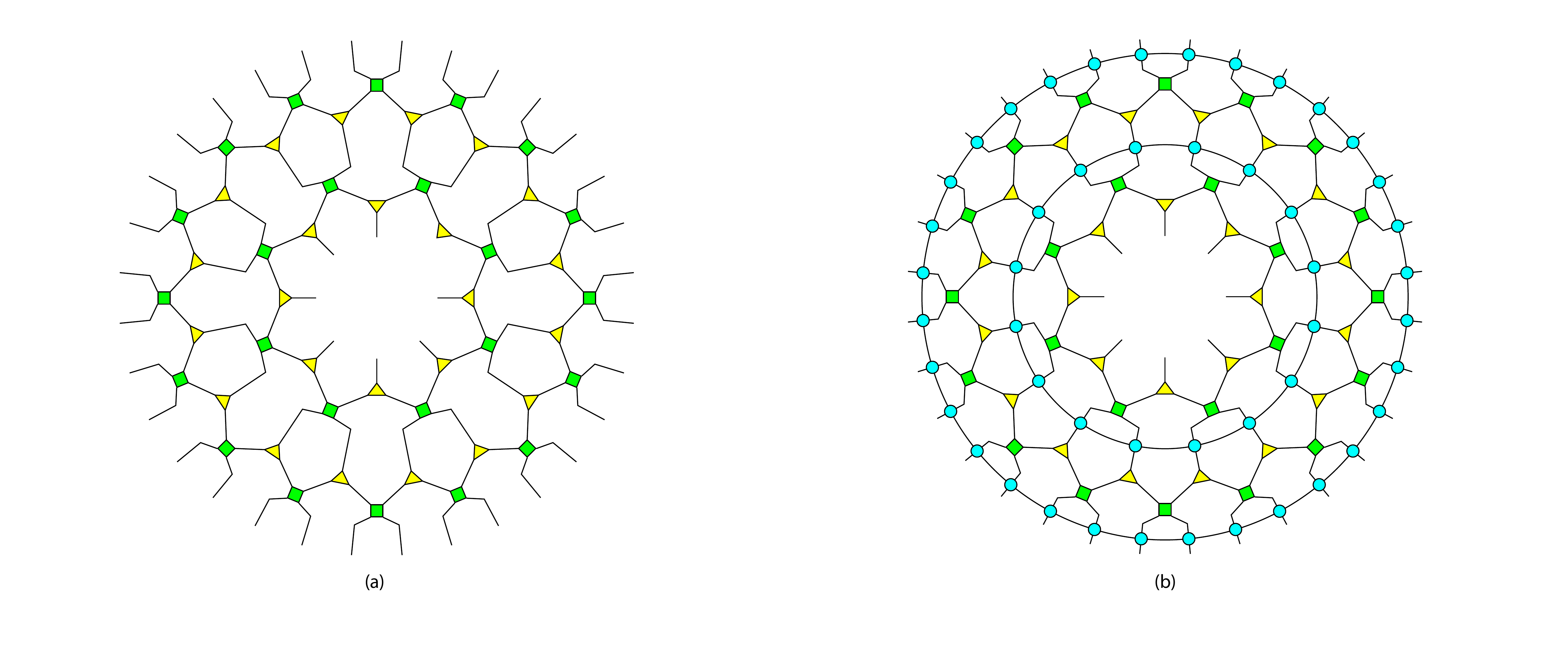}
		\caption{(a) The traditional MERA tensor network on the circle, made of disentanglers $u$ (the green squares) and coarse-grainers $w$ (the yellow triangles).			(b) The Euclidean MERA tensor network on the circle, made of disentanglers $u$ (the green squares) and coarse-grainers $w$ (the yellow triangles), and eclideons $e$ (the blue solid circles).
		}
		\label{fig6}
	\end{center}	
\end{figure}

In this section, we will strictly demonstrate that the tensor network constructed in the previous section can be (approximately) identified with a MERA-like tensor network. It was shown that the tensor network corresponding to the time slice of AdS$_3$ is the one which subtly modifies the traditional MERA tensor network. Explicitly, \cite{Milsted:2018san} proposed a tensor network corresponding to a finite periodic critical quantum spin chain, namely an Euclidean MERA tensor network on the circle, as shown in the FIG.\ref{fig6}(b). The traditional MERA tensor network on the real line (i.e. infinite critical spin chain) is made of infinite layers $\mathcal{W}$ of tensors called disentanglers $u$ and coarse-grainers $w$, while MERA tensor network on the circle is made of finite, periodic layer $\mathcal{W}$ using the same optimized tensors $u$ and $w$, see FIG.\ref{fig6}(a). By interspersing layers $\mathcal{W}$ with layers of euclideons $e$ (tensors that implement euclidean time evolution) \cite{vidal:1412,vidal:1502}, i.e., pre-multiplying $\mathcal{W}$ by a transfer matrix $\mathcal{T}$ made of a row of euclideons $e$ and defining ${\mathcal{W}_+} =\mathcal{W}\mathcal{T}$, one thus obtains the so-called Euclidean MERA tensor network on the circle. A layer $\mathcal{W_+}$ of Euclidean MERA defines a linear map between the Hilbert spaces of the two periodic chains made of $s$ and $s/2$ spins respectively at its boundaries. Then is was argued that the map $\mathcal{W_+}$ matches a path integral on an annulus of the hyperbolic disk, i.e., a time slice of AdS$_3$ and thus assigns this geometry to this kind of MERA-like tensor network~\cite{Milsted:2018san}.

In the following, we are ready to match our tensor network constructed in the previous section with Euclidean MERA tensor network on the circle. The first thing we notice is that our tensor network apparently shows the renormalization process, since from outside to inside, for each layer of operation, we can use the ${\left| \phi  \right\rangle \left| \sigma  \right\rangle }$ states associated with the minimal surface of that layer as ``blocks'', and the ${W^{'}}$ tensor always maps the two  ``blocks'' of the previous layer to a new  ``block'' of the next layer, thus we achieve coarse graining for the boundary state. Meanwhile, the ${W^{'}}$ tensor also realizes the disentangling, because in our construction, the Hilbert space dimension of the ${\left| \phi  \right\rangle \left| \sigma  \right\rangle }$ state associated with the RT surface corresponding to each boundary subregion has been choosen to adapt with the corresponding entanglement entropy, i.e., the logarithm of the dimension exactly equals to the entanglement entropy. According to the subadditivity property of the entanglement entropy, the sum of the entanglement entropy of the two adjacent blocks in the previous layer is larger than the entanglement entropy of the block in the next layer. The exception is the first layer of our tensor network, since in the first layer, $W$ does not represent the two to one map. Actually, it is completely a one to one isometry, which maps the state associated with the ``notch'' (i.e., the RT surface) of the first layer to each boundary subregion reduced state.

However, it is important to note that, in the traditional MERA tensor network, the entanglement entropy $S$ of the boundary subregion is obtained by counting the the number of bonds (denoted as $n$) cut off by the RT surface associated with this region, and then multiplying by the logarithm of the Hilbert space dimension of each bond (denoted as $J$) \cite{Swingle:2009bg,Swingle:2012wq}, i.e.,
\be S \sim n \cdot {\ln{J}}. \ee
While in our construction, it seems that only one bond is associated with the RT surface, whose Hilbert space dimension is denoted as ${\chi _\phi }$, and the entanglement entropy is obtained by
\be S = \ln {\chi _\phi }.\ee
In order for these two construction to be equivalent, we just need the relation
\be{\chi _\phi } = {J^n},\ee
which inspires us to relate the Hilbert space of the ${\left| \phi  \right\rangle \left| \sigma  \right\rangle }$ state associated with the RT surface to the direct product of the Hilbert spaces of the bonds cut off by the RT surface in the traditional MERA tensor network.

To precisely describe this relationship and identify the $W$ tensor with disentanglers $u$ and coarse-grainers $w$, we first absorb the ${\left| \phi  \right\rangle \left| \sigma  \right\rangle }$ state into the ${V^{'}}$ tensor in each step. It's easy to see that this is tantamount to using ${\Psi _V}$ tensors and $W$ tensors in our previous inductive derivation, instead of ${V^{'}}$ and ${W^{'}}$ tensors. Based on this, after the full first layer, we obtain
\be{\left| \Psi  \right\rangle ^{{\rm 1st}}} = {\Psi _V}_N^{{\rm 1st}}\prod\limits_{i = 1}^N {W_i^{{\rm 1st}}}. \ee

And after the second layer of ``surface growth'', we obtain
\be{\left| \Psi  \right\rangle ^{{\rm 2nd}}} = {\Psi _V}_M^{{\rm 2nd}}\prod\limits_{j = 1}^{M = N/2} {W_j^{{\rm 2nd}}} \prod\limits_{i = 1}^N {W_i^{{\rm 1st}}}. \ee

Repeating this process, after the $k$-th layer of ``surface growth'', we have
\be{\left| \Psi  \right\rangle ^{k{\rm th}}} = {\Psi _V}_M^{k{\rm th}}\prod\limits_{i = 1}^{N/{2^{k - 1}}} {\prod\limits_{a = 1}^k {W_i^{a{\rm th}}} }. \ee

Hence we can obtain the representation of FIG.\ref{fig5b}. Note that in FIG.\ref{fig5a} the network is represented by the ${W^{'}}$ tensors, while in FIG.\ref{fig5b} it is represented by $W$ tensors. The subtle difference between the two figures is that, the ${W^{'}}$ tensor (except for those at the first layer representing one-to-one maps) has only inward-pointing legs pointing toward itself, in other words, according to our convention, it has only down-indices, as one can see from eq.(\ref{diff1}), and its legs are not cut off by the RT surfaces explicitly, while the $W$ tensor not only has the down-indices representing the inward-pointing legs pointing to itself from the inner layer, but also the up-indices representing the outward-pointing legs pointing to the outer layer, as can be seem from eq.(\ref{equ8}). Therefore, as shown in FIG.\ref{fig5b}, the arrow directions of legs are both pointing from the center to the boundary in the whole network, we can thus implement a convention that the direction of the coarse-graining map is in the opposite direction of the arrow. Furthermore, now these legs of $W$ tensors are intersecting with the RT surface explicitly, therefore, the informations of ${\left| \phi  \right\rangle \left| \sigma  \right\rangle }$ states are now represented directly by the legs (i.e., bonds) relating the $W$ tensors. In this second representation, it can thus be interpreted as each basis $\left| {\bar m} \right\rangle$ of the state represented by the $\bar \alpha$ bond passing through the RT surface in the OSED tensor network exactly corresponds to the state of an overall configuration in the traditional MERA network, which consists of all the bonds cut off by the RT surface. Therefore, it can be expected that each $W$ tensor should be regarded as a net result of collecting and contracting more than one disentanglers and coarse-grainers in the traditional MERA network, which maps the bonds on two adjacent RT surfaces in the MERA network to the bonds on a larger RT surface.

\begin{figure}[htbp]     \begin{center}
\includegraphics[height=7cm,clip]{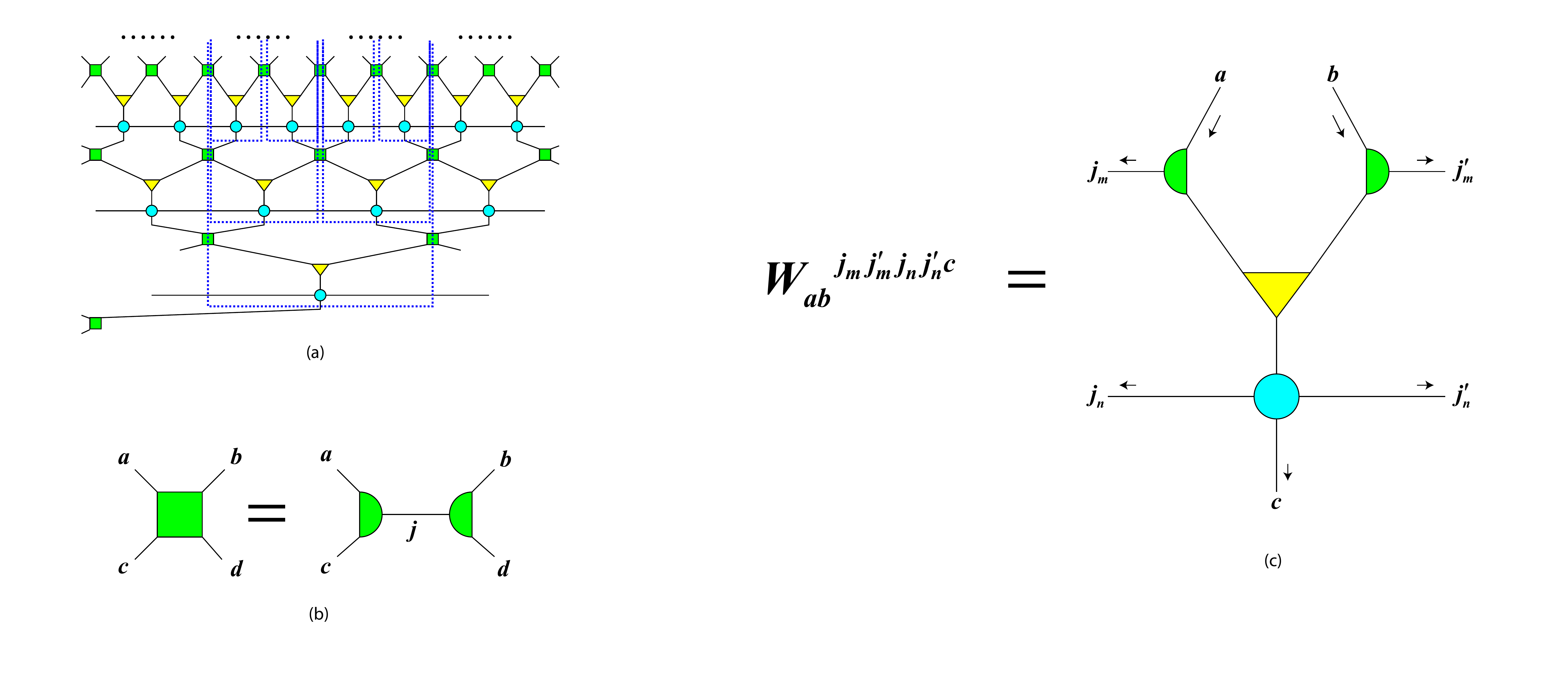}\caption{(a) The RT surfaces (blue curves) in the MERA-like tensor network. (b) The decomposition of disentangler tensor $u$. (c) The corresponding representation of $W$ tensor in MERA-like tensor network.}
\label{fig7}
\end{center}	
\end{figure}

Recall that in the traditional MERA tensor network, roughly speaking, a RT suface is defined as  a curve cutting the minimal number of bonds \cite{Swingle:2009bg,Swingle:2012wq}, which is, more specifically, the boundary of the causal cone of the corresponding boundary subregion \cite{Vidal:2007hda,Vidal:2008zz}, where the causal cone $\mathcal{C}(A)$ of a region $A$ is the set of all sites, unitaries (i.e., disentanglers), and isometries (i.e., coarse-grainers) in the network that can influence the state of $A$. However, for a boundary subregion made of a fairly large number of sites, the width of the causal cone shrinks exponentially as one performs the coarse graining, for a tiny block made of a few sites, the causal cone will actually fluctuate slightly \cite{Swingle:2009bg}. Therefore, in the following matching program, we will determine that the boundary subregions corresponding to the minimal surfaces of the first layer in our tensor network contain a considerable number of lattice sites, and, crucially, we will carefully deal with the definition of RT surface in our tensor network. Our argument is as follows.

Firstly, due to the considerable number of lattice sites, the contribution of the fluctuation of the terminal segment of the causal cone to the entanglement entropy is obviously negligible. Then, considering two adjacent identical boundary subregions ${A_1}$ (on the left) and ${A_2}$ (on the right) which respectively contain $s$ lattie sites ($s$ is a considerable large number) and their union ${A_1}{A_2}$ , we can find that, since according to the definition of causal cone, the outer boundaries of $\mathcal{C}(A_1)$ and $\mathcal{C}(A_2)$ (i.e., the left border of $\mathcal{C}(A_1)$ and the right border of $\mathcal{C}(A_2)$) coincide with borders of $\mathcal{C}({A_1} {A_2})$ respectively, in addition, after a coarse-graining step, the width of $\mathcal{C}({A_1} {A_2})$ shrinks into $s$ sites from $2s$ sites, while the widths of $\mathcal{C}(A_1)$ and $\mathcal{C}(A_2)$ shrink into $s/2$ sites from $s$ sites respectively, and so on, therefore, the inner boundaries of $\mathcal{C}(A_1)$ and $\mathcal{C}(A_2)$ will be always just located at the middle of each layer of $\mathcal{C}({A_1} {A_2})$, that is, these two boudariers coincide. However, strictly speaking, the inner boundaries of $\mathcal{C}(A_1)$ and $\mathcal{C}(A_2)$ are only nearly the same, because each layer's shrinking of causal cones will also have a small fluctuation, which will result in a slight overlap or separation of each layer of the two causal cones. However, also note that our MERA-like tensor network corresponds to a scale invariant critical ground state. It has been demonstrated that scale invariance forces each coarse grained layer to be identical, which leads to that the disentanglers $u$, coarse-grainers $w$, and the additional euclideons $e$ are the same throughout this MERA-like tensor network \cite{Vidal:2007hda,Swingle:2009bg,evenbly:1509}. Therefore, we suggest that in prescribing the RT surface of the MERA-like tensor network, there is actually some kind of freedom to deal with the fluctuations occurring on the causal cone boundaries. We can adopt an effective convention so that the inner boundaries of $\mathcal{C}(A_1)$ and $\mathcal{C}(A_2)$ are exactly the same, as long as the correct entanglement entropy can be obtained according to our convention. In other words, we take the similar view of  \cite{Czech:2015kbp}, which suggested to treat the cut-counting prescription as an empirical fact. According to this viewpoint, we adopt the convention shown in FIG.\ref{fig7}(a) for the causal cone of boundary subregion with a considerable number of lattice sites, where we perform a reasonable decomposition of disentangler tensor $u$ by writing it as a contraction of two $t$ tensors
\be
{u^{abcd}} = {t^{acj}}{t^{bd}}_j,
\ee
where $j$ denotes the auxiliary index that will be contracted, which represents an auxiliary Hilbert space in the tensor network, and we demand that its dimension should be the same as the dimension of the index of the disentangler $u$, i.e., $J$. Thus, in order to correctly calculate the entanglement entropy, it can be viewed that cutting the bonds of disentangles along the causal cone boundary is equivalent to cutting these auxiliary bonds as shown in the figure. Supposing the number of sites contained in the boundary subregion $A$ is $s = \frac{l}{\varepsilon }$, where $l$ is the length of $A$, $\varepsilon$ is the lattice spacing, and using $v$ to denote the number of coarse-graining steps, thus there are $\frac{l}{\varepsilon }{2^{ - v}}$ sites in each corresponding layer of $\mathcal{C}(A)$, when the causal cone ends, we have $\frac{l}{\varepsilon }{2^{ - {v_{\rm end}}}} = 1$, i.e.,
\be{v_{\rm end}} = {\log _2}\frac{l}{\varepsilon }\ee
It can be seen from FIG.\ref{fig7}(a) that, according to our convention, the RT surface will cut off four bonds after each layer, including two auxiliary bonds $j$ and two bonds of euclideons in both sides. Hence, the number of bonds cut off by RT surface is
\be n = 4\int_0^{{v_{end}}} {dv}  = 4{\log _2}\frac{l}{\varepsilon }.
\ee
The entanglement entropy of $A$ is equal to $n$ times $\ln J$, i.e.,
\be {S_A} = n \cdot \ln J = (4{\log _2}J) \cdot \ln \frac{l}{\varepsilon }.
\ee
Note that when one sets
\be\label{centr}\frac {c}{3} = 4{\log _2}J
\ee
the familiar CFT result ${S_A} = \frac{c}{3}\ln \frac{l}{\varepsilon }$ can be recovered, where $c$ is the central charge of CFT. More importantly, the advantage of this approach is that it can be expressed unambiguously that, the inner boundaries of the causal cones of two adjacent equivalent boundary subregions is coincident in the MERA-like tensor network when the fluctuations can be ignored. Now we can determine an explicit expression for the counterpart of the W tensor in the MERA-like tensor network from our tensor network corresponding to the surface growth picture, which can be written in the tensor form specifically as
\be\label{form}
W=W_{ab}^{{j_m}{j_m^{'}}{j_n}{j_n^{'}}c}
\ee
as shown in FIG.\ref{fig7}(c). The reason why $W$ tensor can be represented in such a concise form is mainly because the bonds of the inner boundaries of the two smaller RT surfaces have been contracted, and their outer boundaries share the same indices (except in the last layer) with the larger RT surface. Note that in the figure, the convention of arrow direction has been changed to adapt to the RG flow direction. We thus find the exact counterpart of $W$ tensor in the MERA-like network.

\begin{figure}[htbp]     \begin{center}
\includegraphics[height=7cm,clip]{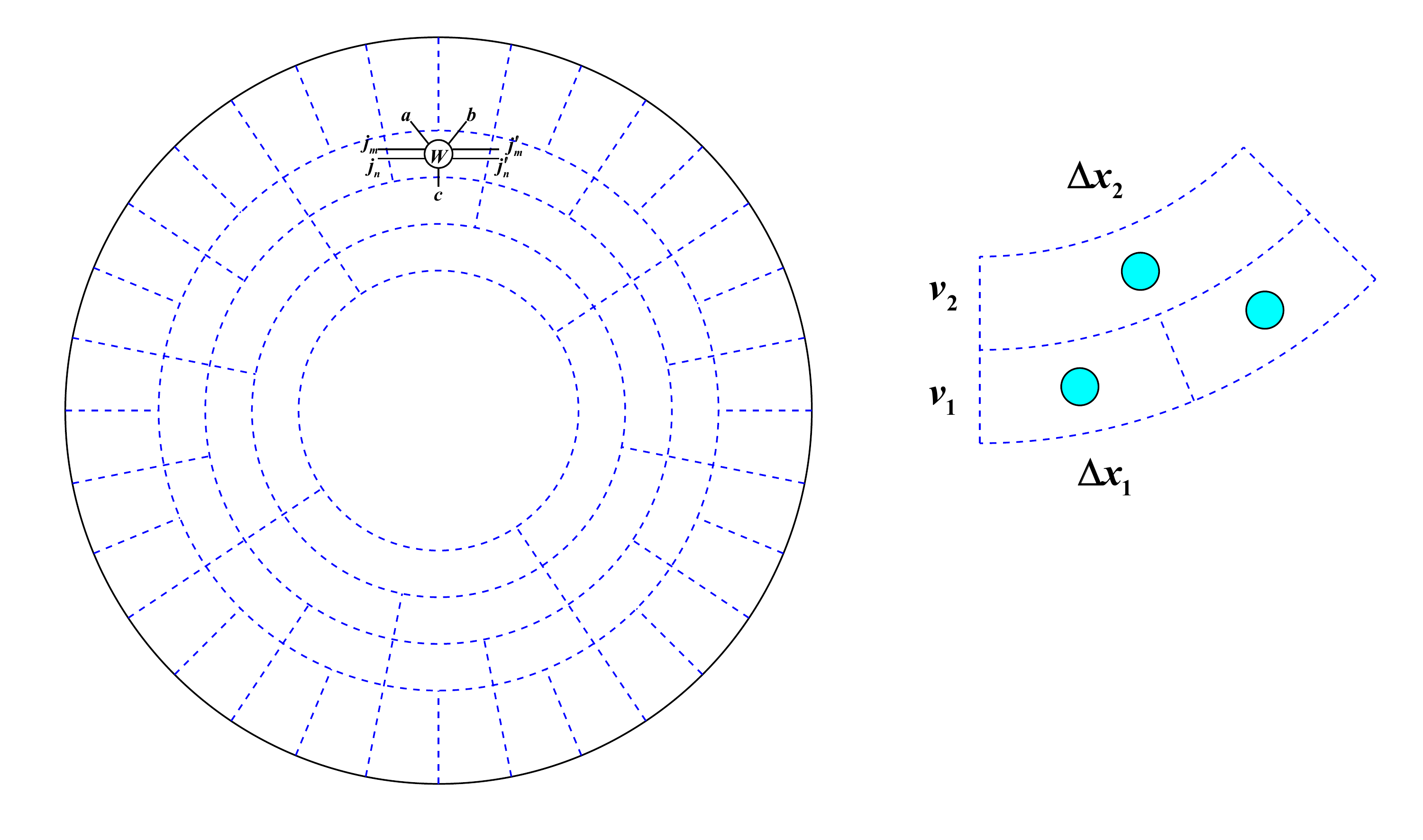}\caption{(a) The MERA-like tensor network is a kind of discretization approximation of our surface growth picture. (b) Each euclideon $e$ can be considered as a cell of bulk spacetime, i.e., the “source” in the ``Huygens’ picture''.}
\label{fig8}
\end{center}	
\end{figure}

Furthermore, due to the fractual property resulting from the iterative process, the relation between the $W$ tensor and the disentangler $u$, coarse-grainer $w$, and euclideon $e$ is the same in each layer, except that the $W$ tensors in in the first layer is interpreted as directly mapping the state on the RT surfaces in the first layer to the corresponding boundary subregions, as we explained previously. Consequently, we have shown that the tensor network that we constructed according to the surface growth scheme and the OSED in the previous sections can be identified with the MERA-like tensor network. More specifically, as shown in FIG.\ref{fig8}(a), the MERA-like tensor network is a kind of discretization approximation of our surface growth picture. This discretization approximation is reflected in that, in the surface growth picture, the RT surfaces corresponding to two adjacent boundary subregions are disjoint, while in the MERA-like tensor network, these two RT surfaces share the same inner boundary. Nevertheless, it has been demonstrated systematically in~\cite{Bao:2018pvs} that a tensor network constructed using the OSED method can be considered as a discretization of the spacetime geometry. Therefore, we can conclude that as long as we define the combination of the specific tensors in the MERA-like tensor network just as in FIG.\ref{fig7}(c) as the formal $W$ tensor in the systematical method, we have obtained a proof that the MERA-like tensor network is indeed a discretized version of the time slice of AdS spacetime.

Our strategy in this section can also be regarded as investigating the MERA-like tensor network from a new point of view, that is, dividing the MERA-like tensor network according to a symmetric pattern with the discrete version of RT surfaces. Interestingly, such a perspective leads to a very intuitive description of an idea proposed in \cite{Milsted:2018yur}, which proposed a $rule~2$ which states that the proper distance between any two nearest neighbor sites in each layer of the tensor network is a constant ${a_{UV}}$, leading to a picture wherein each euclideon $e$ is considered as representing (a path integral on) a square patch of spacetime of size ${a_{UV}} \times {a_{UV}}$, while disentanglers $u$ and coarse-grainers $w$ play the role of gluing together these square pieces into the intended spacetime instead of representing the patches of geometry per se. It can be seen that from our perspective, it is naturally and intuitively to treat each euclideon $e$ as the cell of each small bulk subregion bounded by three RT surfaces, in other words, the ``source'' in our Huygens' picture. To see this more clearly, let us transform the $z$ coordinate to the $v$ coordinate corresponding to the number of coarse-graining steps by $z = {2^v}$, then the time slice of eq.(\ref{Poincare}) becomes
\be\label{L} d{s^2} = {L^2}\left[ {{{\left( {\ln 2} \right)}^2} \cdot d{v^2} + {{\left( {{2^{ - v}}} \right)}^2} \cdot d{x^2}} \right]
\ee
Obviously, the metric in the radial direction is a constant in this coordinate frame, since now the $v$ coordintate directly denotes the number of coarse-graining steps. Thus the size of each small cell is equal in the radial direction. As for another direction, see FIG.\ref{fig8}(b). Because the coordinate length of each circle is the same, according to our discretized picture, we have
\be\label{xv}
\Delta {x_2} &=& 2\Delta {x_1},\\
{v_2} &=& {v_1} + 1.
\ee
The sizes of each cell in the $x$ direction in the outer layer and inner layer are respectively
\be
{a_1} = L \cdot {2^{ - {v_2}}}\Delta {x_2},\\
{a_2} = L \cdot {2^{ - {v_1}}}\Delta {x_1}.
\ee
By eq.(\ref{xv}), we obtain ${a_1} = {a_2}$. So the size of each small cell is also equal in the $x$ direction. Thus it is satisfactory to see that all ideas are consistent.

It is appropriate here to discuss how our work can reconcile the well-known limitations in the AdS/MERA proposal~\cite{Bao:2015uaa}. As one can see from eq.(\ref{L}) that the proper length between adjacent sites in the MERA lattice is of the AdS scale $L$, and it has been shown in~\cite{Bao:2015uaa} that no allowed change of coordinates can lead to sub-AdS scale resolution. Therefore, it is widely believed that the MERA describes geometry only on scales larger than the AdS radius. However, this fact does not affect the potentials of the OSED tensor network as a valid method for describing our surface growth scheme to detect the sub-AdS information. Essentially speaking, the OSED tensor network is more robust and powerful than the MERA-like tensor network in realizing the AdS/CFT correspondence. Note that the OSED tensor network is discretized at a given AdS spacetime, thus in the first layer of discretization of this spacetime (see FIG.\ref{fig2b}), we have the freedom to choose the size of the number $N$ of identical small pieces on the boundary, or equivalently, the size of the length scale of each piece. Therefore, on the one hand, in order to use the OSED tensor network to identify with the MERA-like tensor network to prove its holographic nature, we can choose that each small piece contains a considerable number of lattice sites, although the proper distance between this MERA sites itself is of the order of the AdS radius $L$. Of course, in this situation, this equivalent OSED tensor network faces the same limitation of detecting the sub-AdS information. While on the other hand, we can also go further and do better beyond the MERA-like tensor network, choosing every small boundary subregion piece of the OSED tensor network to be much smaller than the AdS radius, as long as it is still much larger than the Planck scale. In this way one can actually obtain a more refined discretization scheme of spacetime than the MERA-like scheme, although in this case it is open to answer how to find the specific expressions of the $W$ tensor like eq.(\ref{form}), which we will leave it for the future study.\footnote{There are two additional restrictions in the AdS/MERA correspondence~\cite{Bao:2015uaa}, one is that the central charge of the CFT constrains the bond dimension, which is consistent with eq.(\ref{centr}) in the present paper. The other restriction can be roughly described as that the MERA geometry as a real bulk spacetime cannot satisfy the Bousso bound~\cite{Bousso:2014sda,Bousso:2014uxa}. However, although the MERA-like tensor network used in the present paper still faces the same situation, we think this may be due the fact that the MERA-like tensor network is an approximation of the OSED tensor network, which is really dual to the AdS spacetime. It is expected that the OSED tensor network will satisfy the Bousso bound, this requires further examination in the further work. Nevertheless, there is also some strong and reliable argument showing that the improved Euclidean MERA tensor network can faithfully approximate the CFT path integral on a time slice of the AdS$_3$~\cite{Milsted:2018san}, which is consistent with our demonstration.}

\section{More general surface growth scheme}\label{sec-sg}
In this section, we will show that, with the help of the surface/state correspondence, the OSED method above can be generalized to more general surface growth scheme. The key point is that the $W$ tensor is an isometry. Let's still focus on the case of the MERA-like tensor network, which will serve as an illuminating example. As shown in the FIG.\ref{fig8}(a), in order to apply the idea of surface/state correspondence, we can introduce a surface $\Sigma$, which is the envelope of the surface of the first layer (or any layer) and intersects with the bonds  corresponding to indices $a$ and $b$ of the $W$ tensor. Now, if we regard the union of bonds denoted by $a, b$ in the figure as a subregion surface ${\Sigma _A}$ of the surface $\Sigma$, then the union of bonds denoted by ${j_m}j_m^{'}{j_n}j_n^{'}c$ is exactly the corresponding extremal surface $\gamma _A^\Sigma$ anchored on its boundary. According to the idea of surface/state correspondence, the surface $ab$ and surface ${j_m}j_m^{'}{j_n}j_n^{'}c$ then correspond to the state described by the density matrices $\rho \left( {ab} \right)$ and $\rho \left( {{j_m}j_m^{'}{j_n}j_n^{'}c} \right)$ respectively. Now, remarkably, since in our OSED scheme, the $W$ tensor is an isometry, we have
\be\rho \left( {ab} \right)=\rho \left( {{j_m}j_m^{'}{j_n}j_n^{'}c} \right).
\ee
Therefore, we obtain immediately that the von Neumann entropy ${S_{ab}}$ of $ab$ or equally the entanglement entropy $S_A^\Sigma$ (with respect to the region $\Sigma$) is equal to the von Neumann entropy ${S_{{j_m}j_m^{'}{j_n}j_n^{'}c}}$ of ${j_m}j_m^{'}{j_n}j_n^{'}c$,
\be\label{gene} S_A^\Sigma  \equiv {S_{ab}} = S_{{j_m}j_m^{'}{j_n}j_n^{'}c}.
\ee
Note that in the surface/state correspondence, there is a general rule for the extremal surfaces as follows \cite{Miyaji:2015yva}.

$Rule$: The density matrix corresponding to an extremal surface is a direct product of density matrices at each point, which means that the von Neumann entropy of the density matrix of an extremal surface is equal to its area (or in the language of tensor networks, it is just proportional to the number of bonds passing through the surface ).

Thus, actually the above formula eq.(\ref{gene}) is exactly consistent with the generalized RT formula eq.(\ref{gene0}). Note that this argument does not rely on the special choice of the surface growth picture corresponding to the MERA-like tensor network. If one first assume that the surface/state correspondence and generalized RT formula is correct, one can further generalize the OSED proceduce proposed in \cite{Bao:2018pvs} to match our more general surface growth scheme, i.e., the extremal surfaces can be grown out from the arbitrary points of the previous extremal surfaces instehad of the particular points very close to the their end.

To generalize this idea, we first reinvestigate OSED in a more physical sense in the framework of surface/state correspondence. We first accept the general rule mentioned above for any extremal surface ${\Sigma _{\rm ext}}$ in this framework, according to which, there is actually no quantum entanglement within ${\Sigma _{\rm ext}}$, and the von Neumann entropy of the ${\Sigma _{\rm ext}}$ comes totally from the entanglement between the extremal surface ${\Sigma _{\rm ext}}$ itself and its complement in a larger closed surface $\Sigma$ containing ${\Sigma _{\rm ext}}$. The von Neumann entropy of an extremal surface thus behaves as a ``classical'' extensive quantity. The situation is just like that an extremal surface is covered with independent, non-interacting ``coins'' (by ``coins'', we mean ``classical bits''), therefore, we can assign a state $\left| {{\Omega _i}} \right\rangle$ weighted with equal probability ${p_i} = \frac{1}{{{e^S}}}$ to each ``classical'' overall configuration, and thus obtain a mixed state describing this extremal surface, denoted as $\{ (\left| {{\Omega _i}} \right\rangle ,{p_i} = \frac{1}{{{e^S}}})\}$ (i.e., a microcanonical ensemble). It turns out that this understanding can be reconciled with the OSED scheme. In the above OSED prescription, if we ignore the quantum fluctuation effect characterized by state $\left| \sigma  \right\rangle$, each RT surface is described by bond ${\bar \alpha }$ of $\left| \phi  \right\rangle  = \sum\limits_{m = 0}^{{e^{S - O(\sqrt S )}}} {{{\left| {m\bar m} \right\rangle }_{\alpha \bar \alpha }}}$ (i.e., tensor ${\phi ^{\alpha \bar \alpha }}$) intersecting with the RT surface (notice that the other bond $\alpha$ has been absorted into the tensor ${\Psi _V}$ in the previous section according to our convention). The state $\left| \phi  \right\rangle  = \sum\limits_{m = 0}^{{e^{S - O(\sqrt S )}}} {{{\left| {m\bar m} \right\rangle }_{\alpha \bar \alpha }}}$ itself is a pure state (up to a normalized factor), in which the probability amplitude of each basic state ${{{\left| {m\bar m} \right\rangle }_{\alpha \bar \alpha }}}$ is equal, however, since this state is maximally entangled, we can assign a mixed state $\{ (\left| {\bar m} \right\rangle ,{p_m} = \frac{1}{{{e^S}}})\}$ with equal probabilities to the bond $\bar \alpha$. Therefore, each basis $\left| {\bar m} \right\rangle$ of the state corresponding to the $\bar \alpha$ bond passing through the RT surface in the OSED tensor network can be considered as a corresponding ``classical'' overall configuration $\left| {{\Omega _i}} \right\rangle$ on the RT surface, just as we have done in identifying the MERA-like tensor network. With this kind of identification in mind, we can thus understand the OSED operation on this subregion state in the following way: for a subregion $A$ in a pure state of a holographic CFT, whose density matrix characterizes a mixed state $\{ ({\left| i \right\rangle _A},{p_i})\} $ describing this subregion, we first perform a so-called smooth operation on it, as shown by the formula (\ref{formula1}), which shapes the original mixed state $\{ ({\left| i \right\rangle _A},{p_i})\} $ of the subsystem into a particular smoothed state $\{ ({\left| {n\Delta  + m} \right\rangle _A},{{\tilde p}^{\rm avg}}_{n\Delta })\} $. Then, ignoring the non-classical fluctuation, we take each set of basic states (of order $O({e^{\sqrt S }})$) with the same value of $m$ as a group, and map it to a corresponding ``classical'' overall configuration ${\left| {\bar m} \right\rangle _{\bar \alpha }}$ on the RT surface, i.e.,
\be\{ {\left| {\Delta  + m} \right\rangle _A},{\left| {2\Delta  + m} \right\rangle _A}, \cdots ,{\left| {{e^{\sqrt S }}\Delta  + m} \right\rangle _A}\}  \to {\left| {\bar m} \right\rangle _{\bar \alpha }}.\ee
Notice that in this process, the $W$ tensor itself is one-to-one map, since it is also associated with the neglected quantum fluctuation $\left| \sigma  \right\rangle$, i.e.,
\be W:{\left| {n\Delta  + m} \right\rangle _A} \leftrightarrow {\left| n \right\rangle _{\bar \beta}}{\left| m \right\rangle _{\bar \alpha }}.\ee

\begin{figure}[htbp]     \begin{center}
\includegraphics[height=7cm,clip]{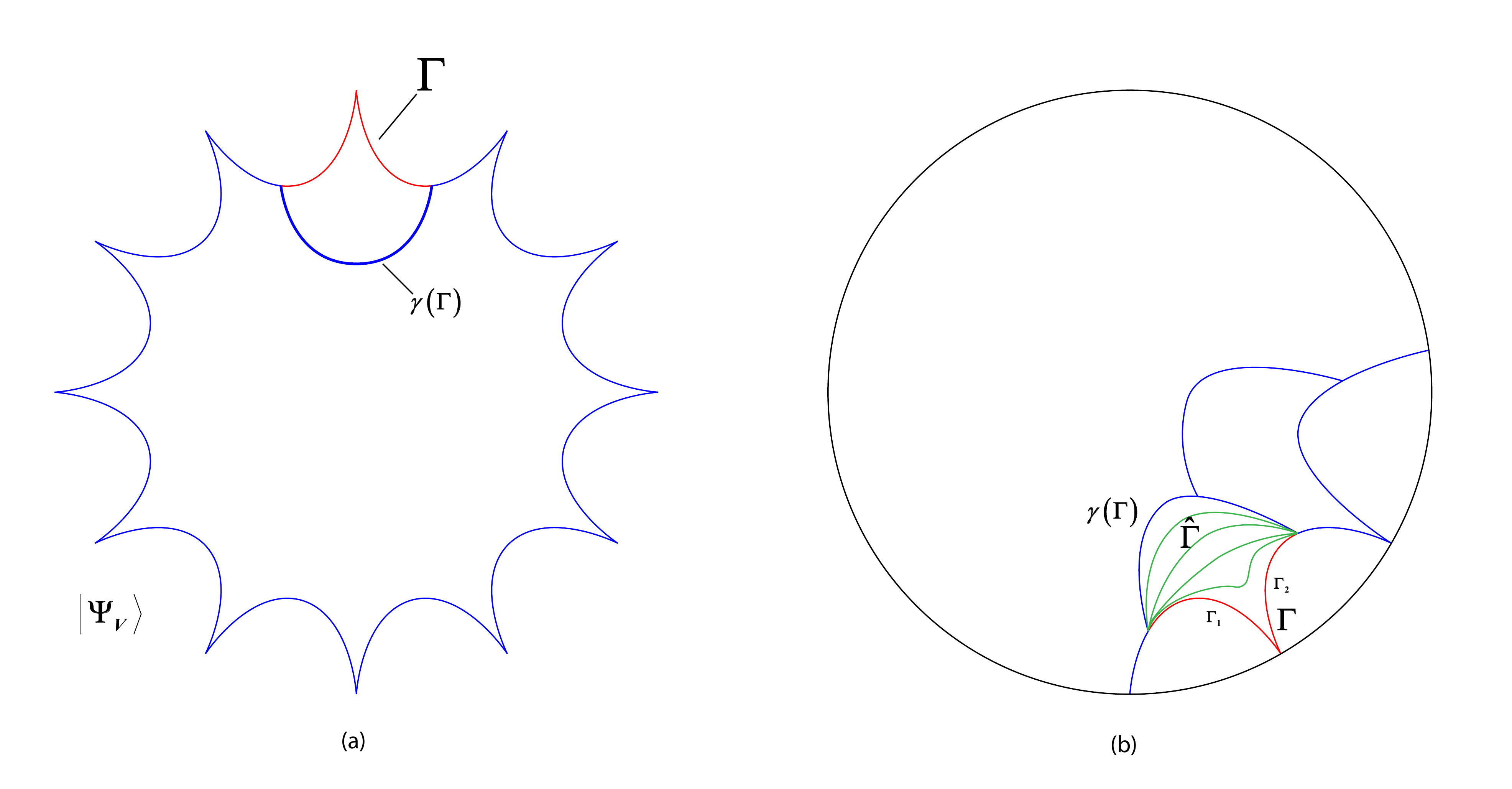}\caption{(a) An example to illustrate the idea of generalized OSED method. (b) More general surface growth scheme.}
\label{fig9}
\end{center}	
\end{figure}

In a word, in the framework of surface/state correspondence we can physically describe the OSED process as follows: we first perform a ``smooth'' operation on the mixed state which describes the subregion information, then by regrouping these smoothed basis states, we holographically map them into a mixed state with equal probabilities on a special extremal surface in a classical geometry, up to some quantum fluctuation. In the same spirit, we can now generalize the OSED method for more general surface growth scheme. For clarity, we give an example to illustrate this idea. We first use the traditional OSED method to construct the first layer as before, which is composed of $n$ equivalent extremal surfaces and their corresponding $W$ tensors, and obtain a state ${\Psi _V}$ of the remaining region that has been cut off $n$ small pieces. Now, in the framework of surface/state correspondence, we have two more weapons: (a) As shown in the FIG.\ref{fig9}(a), if we divide the ``notch'' associated with $\left| {{\Psi _V}} \right\rangle$ state into two parts, denoted as $\Gamma$ and $\bar \Gamma $ respectively, where $\Gamma$ is the union of two segments of different extremal surfaces, then we can define a mixed state ${\psi _\Gamma }$, or equally a density matrix $\rho \left( \Gamma  \right)$ for $\Gamma$ as we did for the boundary subregion. Intuitively, ${\psi _\Gamma }$ can be considered to be defined by the bonds passing through $\Gamma$ in an imaginary or real more general tensor network. (b) If we consider a extremal surface $\gamma \left( \Gamma  \right)$ anchored on the boundary of $\Gamma$, then the generalized RT formula (\ref{gene0}) can assign a physical meaning of entropy to the area of $\gamma \left( \Gamma  \right)$, so that we can assign a Hilbert space dimension matching the value of this entropy to $\gamma \left( \Gamma  \right)$, just as we did for RT surface previously. Consequently, we can generalize the OSED proceduce, i.e., we can not only distill out the $\left| \phi  \right\rangle$ state matching the RT entanglement entropy, but also distill out the $\left| \phi  \right\rangle$ state matching the generalized RT entanglement entropy.

More specifically, similarly to the previous procedure, we perform a ``smooth'' operation on ${\Psi _V}$ state and decompose it, and we obtain a smoothed state $\rho _{\left( \Gamma  \right)}^\varepsilon $ with high fidelity, which satisfies,
\be
{\rm rank}\left( {\rho _\Gamma ^\varepsilon } \right) = {e^{{S_\Gamma } + O\left( {\sqrt {{S_\Gamma }} } \right)}},\nno\\
{p _{\max }}\left( {\rho _\Gamma ^\varepsilon } \right) = {e^{ - {S_\Gamma } + O\left( {\sqrt {{S_\Gamma }} } \right)}}.
\ee
where ${p _{\max }}\left( {\rho _\Gamma ^\varepsilon } \right)$ is the largest eigenvalue of $\rho _{\left( \Gamma  \right)}^\varepsilon $, and ${S_\Gamma }$ is the entropy of the extremal surface $\gamma \left( \Gamma  \right)$ anchored on the boundary of $\Gamma$. Thus the original state ${\Psi _V}$ can be expressed approximately as
\be
\left| {{\Psi _V}} \right\rangle  = \sum\limits_n {\sqrt {{{\tilde p}_n}} {{\left| n \right\rangle }_\Gamma }} {\left| n \right\rangle _{\bar \Gamma }},
\ee
where ${\left| n \right\rangle _\Gamma }$ and ${\left| n \right\rangle _{\bar \Gamma }}$ are the eigenstates of $\Gamma$ and $\bar \Gamma $ respectively, and $\left\{ {{{\tilde p}_n}} \right\}$ are the eigenvalues of the smoothed state $\rho _{\left( \Gamma  \right)}^\varepsilon$. Next, similarly, we rearrange these basis eigenstates such that their probability spectrum ${\tilde p_n}$ is monotonically decreasing and break the resulting sum into blocks of size $\Delta$, and further obtain
\be
\left| {{\Psi _V}} \right\rangle  = \sum\limits_{n = 0}^{{\rm rank}[\rho _\Gamma ^\varepsilon ]/\Delta  - 1} {\sum\limits_{m = 0}^{\Delta  - 1} {\sqrt {{{\tilde p}_{n\Delta  + m}}} {{\left| {n\Delta  + m} \right\rangle }_\Gamma }{{\left| {n\Delta  + m} \right\rangle }_{\bar \Gamma }}} }.
\ee
However, note that now we can choose the size of each block $\Delta$ to exactly match the value of the entropy ${S_\Gamma }$ of extremal surface $\gamma \left( \Gamma  \right)$, or equivalently, the area of $\gamma \left( \Gamma  \right)$ (we can set the proportional factor as 1 for convenience), i.e.,
\be
\Delta  = {e^{{S_\Gamma } - O(\sqrt {{S_\Gamma }} )}}.
\ee
Similarly, we can use the average of the eigenvalues of each block ${\tilde p^{\rm avg}}_{n\Delta }$ to approximate the eigenvalues of the states in each block, and further obtain
\be
\left| {{\Psi _V}} \right\rangle  = \sum\limits_{n = 0}^{{e^{O(\sqrt {{S_\Gamma }} )}}} {\sum\limits_{m = 0}^{{e^{{S_\Gamma } - O(\sqrt {{S_\Gamma }} )}}} {\sqrt {{{\tilde p}^{\rm avg}}_{n\Delta }} {{\left| {n\Delta  + m} \right\rangle }_\Gamma }{{\left| {n\Delta  + m} \right\rangle }_{\bar \Gamma }}} }.
 \ee
Consequently, we can similarly assign a $\left| {{\phi _\Gamma }} \right\rangle\left| {{\sigma _\Gamma }} \right\rangle$ state to the extremal surface $\gamma \left( \Gamma  \right)$ such that the dimension of $\left| {{\phi _\Gamma }} \right\rangle$ can match ${S_\Gamma }$ exactly, while $\left| {{\sigma _\Gamma }} \right\rangle$ will be interpreted as the quantum fluctuation of $\left| {{\phi _\Gamma }} \right\rangle$.
\be
\left| {{\phi _\Gamma }} \right\rangle  &=& \sum\limits_{m = 0}^{{e^{{S_\Gamma } - O(\sqrt {{S_\Gamma }} )}}} {{{\left| {m\bar m} \right\rangle }_{\alpha \bar \alpha }}},\\
\left| {{\sigma _\Gamma }} \right\rangle  &=& \sum\limits_{n = 0}^{{e^{O(\sqrt {{S_\Gamma }} )}}} {\sqrt {\tilde p_{n\Delta }^{avg}} {{\left| {n\bar n} \right\rangle }_{\beta\bar \beta}}}.
\ee
Next, we will construct the isometry tensors $W$ and $V$. Note that in the present case the $W$ tensor maps the state on $\Gamma$ to $\gamma \left( \Gamma  \right)$
\be
W:{H_{\bar \beta}} \otimes {H_{\bar \alpha }} \leftrightarrow {H_\Gamma },
\ee
namely,
\be
W{\left| n \right\rangle _{\bar \beta}}{\left| m \right\rangle _{\bar \alpha }} = {\left| {n\Delta  + m} \right\rangle _\Gamma },
\ee
and for $V$, we have
\be
V:{H_\beta} \otimes {H_\alpha } \leftrightarrow {H_{\bar \Gamma }},
\ee
namely,
\be
V{\left| n \right\rangle _\beta}{\left| m \right\rangle _\alpha } = {\left| {n\Delta  + m} \right\rangle _{\bar \Gamma }}.
\ee
Thus, we obtain
\be
{\Psi _V} = W_{\bar \beta\bar \alpha }^\Gamma V_{\beta\alpha }^{\bar \Gamma }{\phi ^{\alpha \bar \alpha }}{\sigma ^{\beta\bar \beta}} = W_{\bar \beta\bar \alpha }^\Gamma {\left( {V\left| \phi  \right\rangle \left| \sigma  \right\rangle } \right)^{\bar \Gamma \bar \alpha \bar \beta}},
\ee
and the original state corresponding to the whole spacetime can be written as
\be
\Psi  = {\left( {{W^{{\rm 1st}}}} \right)^N}W_{\bar \beta\bar \alpha }^\Gamma {\left( {V\left| \phi  \right\rangle \left| \sigma  \right\rangle } \right)^{\bar \Gamma \bar \alpha \bar \beta}}.
\ee
Until now, we have explicitly accomplished the generalized OSED procedure. Similarly, we can further define the ${W'}$ tensor and ${V'}$ tensor as before, so that all the $\left| {{\phi _\Gamma }} \right\rangle\left| {{\sigma _\Gamma }} \right\rangle$ state associated with the extremal surfaces involved in the discretization scheme can be clearly shown in $\Psi$. However, we prefer to take the viewpoint we adopted in Sec.\ref{sec-tnl}, i.e., using the bonds intersecting with the minimal surfaces to represent the states associated with the them directly. Obviously, the above procedure does not depend on our special discretization scheme of the first layer. Therefore, as long as the surface/state correspondence and the generalized RT formula are valid, this approach can be applied to any surface growth scheme, as shown in FIG.\ref{fig9}(b), and hopefully, even can be applied to more general spacetimes.

Based on the above analysis, we argue that the surfaces growth process is in fact a kind of OSED process. To see this more explicitly, considering a new extremal surface grows from the endpoints of the ${\Gamma _1}$ part and ${\Gamma _2}$ part of two adjacent extremal surfaces, as shown in FIG.\ref{fig9}(b), there is no internal quantum entanglement within ${\Gamma _1}$ or ${\Gamma _2}$ itself, however, when we consider their union as a whole $\Gamma$, there exists the mutual entanglement between the two (which essentially comes from the internal entanglement within the boundary subregion $A$ supporting the initial growing of extremal surfaces), therefore, we can use a mix state ${\psi _\Gamma }$ to describe the whole $\Gamma$. The process of growing this new extremal surface can then be regarded as ``propagating'' the information of the mix state ${\psi _\Gamma }$ to the new extremal surface with high fidelity. Note that since this information will be ``decoded'' by calculating the area of the new extremal surface in a classical sense, thus before the propagation, it is necessary to ``encode'' ${\psi _\Gamma }$ by a ``classical smooth'' operation. Intuitively, when the quantum correction is neglected, the ${\psi _\Gamma }$ can be encoded by this operation into a delicate mixed state, in which each basis state is weighted with equal probability, and thus can correspond to a ``classical configuration'' ${\left| {\bar m} \right\rangle _{\bar \alpha }}$ on the new extremal surface. In this way, the surface growth process successfully maps (in other words, ``propagates'') the ${\psi _\Gamma }$ state to a mixed state with equal probabilities on a special extremal surface $\gamma \left( \Gamma  \right)$ in the classical geometry up to some negligible quantum fluctuation, holographically. Since the entanglement between ${\Gamma _1}$ and ${\Gamma _2}$ comes from the internal entanglement in the initial boundary subregion $A$, while all the growing extremal surfaces is contained within the entanglement wedge of $A$, we can conclude that the new extremal surface $\gamma \left( \Gamma  \right)$ within the entanglement wedge of the boundary subregion $A$ can indeed detect the information inside $A$.

From the perspective of the tensor network, this kind of surface growth process can always be characterized by the isometry tensor $W$, which plays the role of ``distilling'' the state on some regions of the previous extremal surfaces to the new grown extremal surfaces. For example, in FIG.\ref{fig9}, the $W$ tensor distills the state of $\Gamma$ to the extremal surface $\gamma \left( \Gamma  \right)$. It turns out that each general surface growth scheme actually corresponds to a tensor network that can both approximate the boundary state and match the geometry of bulk spacetime. We thus claim that, with the help of the surface/state correspondence and generalized RT formula, we further generalized the OSED tensor network proposed by \cite{Bao:2018pvs}. In addition, it is quite satisfactory to see that our work is naturally reconciled with a claim in \cite{Miyaji:2015yva}, which suggests that if one considers a serious of surfaces obtained by performing a continuous smooth deformation on some convex surface with its two boundaries fixed (until it reaches a extremal surface), then the density matrices of these surfaces should be related by the unitary transformation, therefore the von Neumann entropy of the deformed surface will not change under this deformation, and in particular, match the area of the extremal surface exactly when the deformed surface reaches the extremal surface. According to this claim, we can imagine a surface $\hat \Gamma $ slowly deformed by the extremal surface $\gamma (\Gamma )$ and make it closer and closer to the $\Gamma$ surface, as shown in FIG.\ref{fig9}(b), and thus reach a conclusion that the von Neumann entropy of the density matrix of $\Gamma$ is equal to the entropy of $\gamma (\Gamma )$, which is consistent with our generalized OSED scheme, or equivalently, the surface growth scheme.

\section{Conclusions and discussions}\label{sect:conclusion}
In this paper, inspired by Huygens' principle of wave propagation, and based on the surface/state correspondence, we proposed an intriguing surface growth scheme for the bulk reconstruction. Moreover, our surface growth picture naturally relates to a corresponding OSED tensor network, which is constructed by using and extending the OSED method. The characteristic of this kind of tensor network is that the dimension of Hilbert space of the state on RT surface matches exactly the entanglement entropy of its corresponding boundary subregion. According to a special surface growth picture with spherical symmetry and fractal feature, as shown in FIG.\ref{fig9}(b), we first constructed a corresponding OSED tensor network mainly characterized by the $W$ tensor, which maps the states of two adjacent RT surfaces into a larger RT surface containing the former. Interestingly, we showed that, by a careful treatment of the definition of RT surfaces, in which the inner boundaries of the RT surfaces associated with two adjacent equivalent boundary subregions are taken as to coincide, this tensor network can be identified with the MERA-like tensor network. In this manner, we successfully identified the tensor network constructed by us with the MERA-like tensor network, and accordingly concluded that as long as we define the combination of the MERA tensors in a specific way (approximately) as the formal $W$ tensor in that systematical method, we have obtained a proof that the MERA-like tensor network is indeed a discretized version of the time slice of AdS spacetime, rather than just an analogy, since it has been proved in \cite{Bao:2018pvs} that the tensor network constructed by this systematical method can reproduce the correct boundary state with high fidelity, and have a bulk geometry that matches the bulk AdS spacetime perfectly. Furthermore, we showed that our perspective is consistent with the idea that each euclideon tensor $e$ can be regarded as a square patch of spacetime \cite{Milsted:2018yur}, and even make it more intuitive.

Moreover, the success of this identification with MERA-like tensor network inspired us to further extend the original OSED method to describe more general surface growth picture. More specifically, with the help of the surface/state correspondence and generalized RT formula, we endowed an explicit physical meaning for bulk extremal surfaces far away from the boundary, that is, we defined for each extremal surface an appropriate mixed state $\{ (\left| {\bar m} \right\rangle ,{p_m} = \frac{1}{{{e^S}}})\}$ which can match the $\bar \alpha $ bond in the OSED tensor network exactly (without taking quantum corrections into account). Then we explicitly presented a procedure similar to \cite{Bao:2018pvs} to show that the entanglement within the two adjacent segments of extremal surfaces in the previous layer can indeed be distilled out to the new grown extremal surface anchored on the endpoints of these two segments.

The generalization of OSED scheme has several important significances. First of all, using this generalized OSED method, we can construct tensor network corresponding to more general surface growth cases. Secondly, this generalization leads to a more profound understanding of the surface growth process. We argue that the process of growing a new extremal surface is actually a kind of classical encoding operation on the entanglement within the previous extremal surfaces. In other words, the information of entanglement is encoded into a delicate mixed state up to some quantum fluctuation, such that when it propagates to the new extremal surface, it can be decoded by calculating the area of this new extremal surface in the classical sense. Furthermore, since the entanglement between the extremal surfaces of each layer comes from the entanglement within the initial boundary subregion, while all these extremal surfaces is contained within the entanglement wedge of the initial boundary subregion, we thus conclude that these grown extremal surfaces within the entanglement wedge can indeed detect the information inside the boundary subregion, which provides an explicit and efficient approach for bulk reconstruction in the entanglement wedge and also consistent with the idea of subregion duality.

\section*{Acknowledgement}
We would like to thank L.-Y. Hung for helpful discussions. J.R.S. was supported by the National Natural Science Foundation of China (No.~11675272). The Project is also funded by China Postdoctoral Science Foundation
(No. 2019M653137)





\end{document}